\documentclass[conference,compsoc]{IEEEtran}
\ifCLASSOPTIONcompsoc
  \usepackage[nocompress]{cite}
\else
  \usepackage{cite}
\fi

\ifCLASSINFOpdf
\else
\fi
\usepackage{setspace}
\usepackage{cite}
\usepackage{amsmath,amssymb,amsfonts}
\usepackage{algorithmic}
\usepackage{graphicx}
\usepackage{textcomp}
\usepackage{xcolor}
\usepackage{multirow}
\usepackage{tcolorbox}
\usepackage{enumitem}
\usepackage{listings}
\usepackage{subcaption}
\usepackage{framed, multido}
\usepackage{amsmath}
\usepackage{amsthm}
\usepackage{booktabs}
\usepackage{algorithm}
\usepackage[switch]{lineno}
\usepackage{tikz}
\usepackage{svg}
\usepackage{diagbox}
\usepackage{makecell}
\usepackage{colortbl}
\usepackage{soul}
\usepackage{ragged2e} 
\usepackage{booktabs,makecell, multirow, tabularx}
\newcommand*\emptycirc[1][1ex]{\tikz\draw (0,0) circle (#1);} 
\newcommand*\halfcirc[1][1ex]{%
	\begin{tikzpicture}
		\draw[fill] (0,0)-- (-90:#1) arc (-90:90:#1) -- cycle ;
		\draw (0,0) circle (#1);
\end{tikzpicture}}
\newcommand*\fullcirc[1][1ex]{\tikz\fill (0,0) circle (#1);}

\usepackage{xcolor}
\usepackage{xspace}

\def\BibTeX{{\rm B\kern-.05em{\sc i\kern-.025em b}\kern-.08em
    T\kern-.1667em\lower.7ex\hbox{E}\kern-.125emX}}

\makeatletter
\newcommand{\linebreakand}{%
  \end{@IEEEauthorhalign}
  \hfill\mbox{}\par
  \mbox{}\hfill\begin{@IEEEauthorhalign}
}
\makeatother

\begin{document}
%
\title{Releasing Malevolence from Benevolence:\\The Menace of Benign Data on Machine Unlearning
}


\author{
\IEEEauthorblockN{
Binhao Ma\IEEEauthorrefmark{1}\IEEEauthorrefmark{2}
Tianhang Zheng\IEEEauthorrefmark{1}\IEEEauthorrefmark{2}
Hongsheng Hu\IEEEauthorrefmark{3},
Di Wang\IEEEauthorrefmark{4}, 
Shuo Wang\IEEEauthorrefmark{5}, 
Zhongjie Ba\IEEEauthorrefmark{1}\IEEEauthorrefmark{2}, \\ 
Zhan Qin \IEEEauthorrefmark{1}\IEEEauthorrefmark{2}, 
Kui Ren \IEEEauthorrefmark{1}\IEEEauthorrefmark{2}
}

\IEEEauthorblockA{\IEEEauthorrefmark{1}The State Key Laboratory of Blockchain and Data Security, Zhejiang University}
\IEEEauthorblockA{\IEEEauthorrefmark{2}Hangzhou High-Tech Zone (Binjiang) Institute of Blockchain and Data Security}
\IEEEauthorblockA{\IEEEauthorrefmark{3}CSIRO’s Data61}
\IEEEauthorblockA{\IEEEauthorrefmark{4}King Abdullah University of Science and Technology}
\IEEEauthorblockA{\IEEEauthorrefmark{5}Shanghai Jiao Tong University}
\IEEEauthorblockA{Corresponding Author: Tianhang Zheng \quad Email: zthzheng@gmail.com}\\\\}











\maketitle


\begin{abstract}

Machine learning models, when trained on massive real or synthetic data, typically can achieve exceptional prediction performance in various application domains. However, the outstanding model utility comes with growing privacy concerns since the data used for model training may contain sensitive information. To mitigate privacy concerns, machine unlearning has been proposed to eliminate the information of specific data samples from machine learning models. 
While some machine unlearning techniques can optimize the models to forget data at a low cost, recent works show that a malicious user could request unlearning on perturbed data to compromise the unlearned model. Despite the effectiveness of the attack, the perturbed data does not match the original data for training the original model and thus cannot pass hash verification. Moreover, the existing attacks on machine unlearning suffer from limited practicality and applicability, due to their demand on additional knowledge and non-negligible attack budgets.

To fill the gaps in the existing unlearning attacks, we propose a new attack called the Unlearning Usability Attack, which is model-agnostic, unlearning-agnostic, and budget-friendly.
An unlearning usability attack is implemented by distilling data distribution information into a small quantity of data, which is labeled as benign data by automatic poisoning detection tools due to its positive contribution to model training. Although the data is benign for machine learning, unlearning on the data will induce significant loss of data information in the models.
Our evaluation reveals that, under different attack scenarios, unlearning the benign data, which is no more than $1\%$ of the entire training data, will drop the model accuracy by up to $50\%$. Our evaluation also indicates that the well-prepared benign data naturally act as hard-to-unlearn samples in recent unlearning techniques, as the process of erasing these synthetic instances demands a higher budget than regular data. These new findings motivate future research to rethink ``data poisoning'' in the context of machine unlearning.

\end{abstract}


%
\IEEEpeerreviewmaketitle


\section{Introduction}


Many breakthroughs in modern machine learning owe their success to the abundance of available data. However, since the data may contain sensitive information of individuals~\cite{r1}, organizations must adhere to data protection regulations during the data collection and utilization processes, prioritizing the privacy of the individuals to minimize the risks of legal action and reputational harm. To give individuals full control of their data, prominent regulations such as General Data Protection Regulation (GDPR)~\cite{r2} and California Consumer Privacy Act (CCPA)~\cite{r3} have legally established the right to ``erasure of personal data without undue delay.'' Notably, Google Search has received millions of individuals' requests to remove specific URLs from search results within a five-year period~\cite{r5}, demonstrating the substantial demand from individuals for stronger privacy protection of their personal data online.


Complying with privacy regulations, the concept of machine unlearning has been proposed and broadly studied in recent literature~\cite{r8,r9,r10,r11}.
Machine unlearning necessitates that upon an individual's request for certain data (i.e., the unlearned data) removal, the machine learning model owner must ensure that the unlearned data is erased from the trained model, safeguarding the information of the unlearned data from potential privacy attacks, including model inversion attacks~\cite{r6}, data extraction attacks~\cite{carlinimembership,carlinisecret} and membership inference attacks~\cite{r7}. 
Retraining the model from scratch without the unlearned data is an intuitive solution for machine unlearning, but it leads to high computational overhead, especially when the model is complex and trained on large-sized datasets. For example, it is almost impossible to retrain a large language model (LLM) like ChatGPT~\cite{achiam2023gpt} when some unauthorized data have to be removed from the model due to copyright issues~\cite{openai}. 

Besides naively retraining for unlearning, approximate unlearning methods~\cite{r8,r9,r10,r12} that directly update the trained model for deleting the unlearned data offer an unlearning alternative with the advantages of efficiency and lightweight. For example, the work~\cite{maini2024tofu} shows that some approximate unlearning methods enable the LLM of Llama-2-7B~\cite{touvron2023llama} to forget the unlearned data as the unlearned model has never trained on it. However, despite the promising practicability, approximate unlearning opens new attack surfaces for attackers, who may intentionally perform malicious activities to compromise the unlearned model during the unlearning process. This is because approximate unlearning involves complex trade-offs between model utility and unlearning effectiveness, and it is difficult to quantify how much an unlearned sample can influence the trained model~\cite{hu2023duty}.


\noindent \textbf{Motivation.} There are several works~\cite{hu2023duty,marchant2022hard,liu2024backdoor,qian2023towards,di2022hidden} starting to investigate the security vulnerability of machine unlearning. Specifically, they assume that a malicious user can upload crafted unlearned data to compromise the unlearned model, e.g., reducing the robustness of the model~\cite{qian2023towards}. 
\ul{However, despite the effort these pioneering works made to reveal the vulnerability of machine unlearning, we identify three gaps in existing attacks for an automatic unlearning pipeline with hash verification.}

\noindent $\bullet$ \textbf{Gap 1: Deny of the Deletion Requests.} Asking for the deletion of the crafted unlearned data involves an assessment of the erasure requests, which may result in a \textit{denial-of-unlearning-service}. This is because the crafted data no longer matches the original data used for training the original model. Despite the work~\cite{hu2023duty} auguring that the model owner should not store the original dataset after training the model, hash verification techniques~\cite{venkatesan2000robust} can be employed by the model owner to reject malicious unlearning requests.

\noindent $\bullet$ \textbf{Gap 2: Limited Practicability and Applicability of the Attack.} The attacks proposed in existing works~\cite{hu2023duty,marchant2022hard,liu2024backdoor,qian2023towards,di2022hidden} require the white-box or black-box access to the target model. In addition, some attacks~\cite{liu2024backdoor,qian2023towards,di2022hidden} also require to know which unlearning method the model owner uses. That is, these attacks are model-specific and unlearning-specific: the crafted unlearning data is only effective for specific models under specific unlearning methods. This limitation heavily restricts the applicability and practicability of such attacks, as sometimes it is difficult for an attacker to obtain the information of the model and the deployed unlearning mechanism, e.g., in commercialized machine learning as a service (MLaaS) scenarios where such information is kept private.

\noindent $\bullet$ \textbf{Gap 3: Unrealistic Amount of Unlearning Samples.} Some of the existing attacks~\cite{qian2023towards} unnecessarily require a large amount of unlearning samples for achieving the threats. For example, to heavily compromise the utility of the unlearned model on a specific class, the attack in~\cite{hu2023duty} requires the model owner to unlearn 50\% well-crafted samples (2,000 samples in the case of CIFAR-10 \cite{r32}) of that class. However, from the perspective of the model owner, receiving such a large amount of unlearning samples may trigger the data owner to be aware of potential threats to the model, restricting the applicability of the attacks.

\noindent \textbf{Contributions.} The gaps in existing works make them inappropriate to fully capture the vulnerability of machine unlearning in practice. To fulfill the identified gaps and better understand the vulnerability of machine unlearning in practice, in this paper, we propose a new attack called \textit{unlearning usability attack} in machine unlearning. Specifically, being \textit{model-agnostic}, \textit{unlearning-agnostic}, \textit{with small amount of unlearning samples}, and \textit{without the need to modify the unlearned data}, the unlearning usability attack anticipates the usability threat of the unlearned model under the MLaaS environment: a malicious user can legalistically leverage his right to be forgotten to crash the unlearned model, making it useless for other normal users (detailed in Section \ref{sec:threatmodel}).

\begin{figure}[t]
\centering
\includegraphics[width=1.0\linewidth]{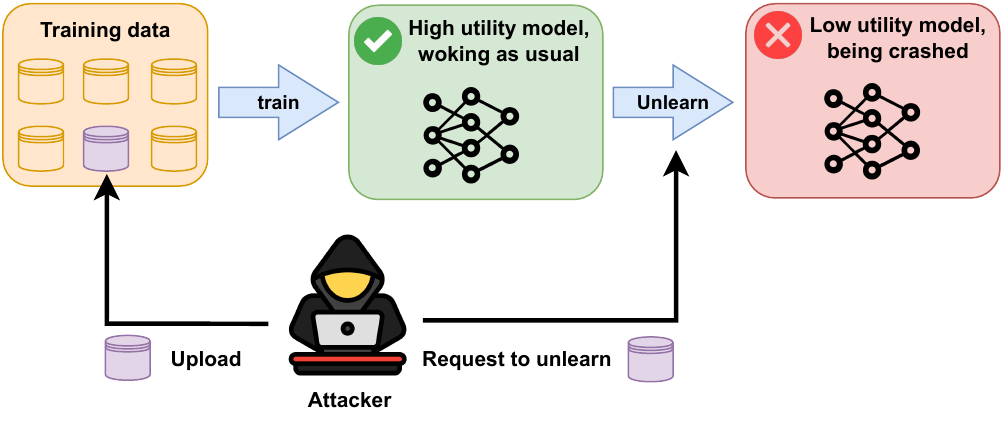}
\caption{An overview of the unlearning usability attack. An attacker first contributes data to train the model and then revokes the contribution to crash the model.}
\label{fig:attacks}
\end{figure}

\noindent \textbf{Unlearning Usability Attacks.} There are two steps for a malicious user to perform the unlearning usability attack: contribute and then revoke, to be short. Specifically, as depicted in Figure~\ref{fig:attacks}, the attacker first contributes his well-prepared data to train the model. Then, the attacker exercises his right to be forgotten for unlearning, revoking his contribution to the trained model. When the unlearning requests are fulfilled, the unlearned model largely reduces its utility, making the resulting unlearned model useless for other users. The attack is practically possible in the scenarios of insider threats, where the insider attacker intends to deliberately harm the business interests of a broader group by misusing their authorized rights.

\noindent \textbf{Attack Intuition.} The intuition of the unlearning usability attack is that by providing well-prepared informative data to promote the training of the original model, the attacker can then revoke his contribution through unlearning to compromise the unlearned model. Because such well-prepared data is highly informative to the original model, the unlearned model inevitably largely loses its utility, as the corresponding knowledge has been deleted. To achieve the attack, we consider the attacker can leverage dataset condensation~\cite{r27} techniques, which can condense the knowledge of many samples into a few informative samples.

\noindent \textbf{Benign Data for Attack (Difference from Poisoning Attacks).} Note that in our attacks, the well-prepared data is different from the poisoning data in traditional poisoning attacks~\cite{rbad,rblend,rlc}. Specifically, in traditional poisoning attacks, the attacker intentionally poisons the training dataset to compromise the model, e.g., creating backdoors into the model or implementing target attacks. The model behaves unmorally because the poisoned data is harmful during the training process. \ul{However, in our attacks, the well-prepared data is \textbf{\em benign} for machine learning since it makes a positive contribution to the model training. Put differently, the data is helpful during the training process, while it only exhibits its menace on the model during the unlearning process.} Therefore, the unlearning usability attack sheds light on a new perspective to rethinking ``poisoning'' in the context of machine unlearning.

Our contributions can be summarized as follows:

\noindent $\bullet$ This paper proposes the unlearning usability attack, identifying an impact vulnerability in machine unlearning. We show that the unlearned model can suddenly crash, i.e., lose its utility when the original model unlearns the deliberately prepared data contributed by the attacker, while such data is helpful for the model during the training process.

\noindent $\bullet$ The proposed unlearning usability attack demonstrates its practicability by being model-agnostic and unlearning-agnostic, i.e., without the knowledge of the target model and the unlearning algorithm. With a small number of well-prepared unlearning samples, e.g., 1\% of the whole dataset, the unlearned model loses 50\% (at least 5 times) accuracy compared to normal unlearning.

\noindent $\bullet$ We conduct extensive experiments on representative unlearning methods, across various model architectures and benchmark datasets, and under various settings. The experimental results validate the effectiveness of the proposed unlearning usability attacks in largely crashing the model's utility. The source code will released on GitHub after the acceptance of this paper.

\section{Related Work}\label{sec:related-work}
\noindent \textbf{Machine Unlearning.} Machine unlearning methods can broadly be divided into two categories: exact unlearning and approximate unlearning. Exact unlearning involves retraining the model from scratch, i.e., retraining on the dataset minus the data slated for removal. While this kind of method requires particular skills to be executed efficiently, its primary advantage is that it guarantees the total elimination of any impact from the unlearned data, as the retrained model has never been trained on that data. The work~\cite{r17} introduced a technique for forgetting points from clustering. The key idea of both methods involves partitioning the data into independent sections and aggregating the final model from sub-models trained on these partitions. The idea is later extended to graph unlearning~\cite{wang2023inductive,chen2022graph}. While effective for forgetting data points by retraining only the affected partitions, this method introduces significant storage overhead and inference latency. Brophy et al.~\cite{brophy2021machine} suggest a forest structure designed for data removal that facilitates the effective unlearning of samples in random forests. Approximate unlearning modifies the parameters of the trained model to mimic a model retrained from scratch, achieving the unlearned state. Typically, approximate unlearning involves updating the trained model's parameters through a limited number of iterations, using data derived from the information to be unlearned. Guo et al.~\cite{r21} presented an unlearning method for linear regression, but its applicability to nonlinear methods is limited. The works~\cite{r8,r9,r10} further addressed these limitations. Graves et al.~\cite{r10} proposed preserving only the parameters updated for unlearned samples, allowing the model to forget the original samples when needed for unlearning. Warnecke et al.~\cite{r12} suggested adding noise to the samples and then making the model forget both noisy and original samples through loss updates, which can effectively erase information related to the samples.

\noindent \textbf{Attacks on Machine Unlearning.} Machine unlearning has yielded promising outcomes, fostering an optimistic outlook for user privacy protection in third-party services. However, it has concurrently opened up new avenues for attackers to exploit model vulnerabilities. Di et al.~\cite{r28} introduced the camouflaged data poisoning attack, where the accuracy of the model's predictions is negatively impacted when the attacker triggers a request to remove a subset of data samples from the dataset. However, implementing this method is challenging as it necessitates an understanding of the targeted network architecture and knowledge of the training procedure, aspects that are often difficult to attain in practical applications. Marchant et al.~\cite{r14} introduced a method that employs projected gradient descent (PGD) to add perturbations to images. These manipulated images are then uploaded to the server to attack linear model unlearning, making it challenging for the model to forget such adversarial examples. Hu et al.~\cite{hu2023duty} proposed optimizing images beyond the decision boundary and subsequently uploading them through MLaaS, inducing excessive unlearning in the model. Nevertheless, both of these attack methods can be thwarted through a straightforward hash comparison~\cite{r16} between the uploaded samples and the original samples. Additionally, the works~\cite{r17,r18} suggested that model unlearning can be achieved solely through the user-provided index, making these attacks unsuccessful in such scenarios.

\noindent \textbf{Difference from Existing Attacks.} The biggest difference between our unlearning usability attacks and existing attacks~\cite{hu2023duty,marchant2022hard} is that the existing attacks allow the attacker to post-revising their unlearned data to compromise the model, while we do not allow that. In our case, the attacker exercises the right to be forgotten by requesting to unlearn exactly what the attacker provided for training the model. In addition, different from existing works~\cite{qian2023towards} that require unlearn a large amount of data samples for compromising the unlearned model, our attacks crash the utility of the model with the usage of only a few samples, e.g., no more than 1\% of the whole training dataset. Last, the closest work to our paper is~\cite{hu2023duty}, which also exploits how an unlearned model can unintentionally reduce its utility through over-unlearning attacks. Our work differs from this work in two aspects. First, the attacker in our setting can not post-modify the unlearning data. Second, the work~\cite{hu2023duty} focused on how the added perturbations in the unlearned data can additionally reduce the utility of the unlearned model, while we focus on how \textit{benign training data} can invalidate the model's utility. As detailed in Table~\ref{tb1}, we summarize and compare our unlearning usability attacks with existing attacks in machine unlearning.

\begin{table}[t]
	\centering
	\caption{Comparison of our unlearning usability attacks and existing attacks in machine unlearning. \fullcirc\; represents available and \emptycirc\; denotes unavailable.}
	\resizebox{0.5\textwidth}{!}{
	\begin{tabular}{lcccccc}
		\toprule
		 & \multicolumn{2}{c}{\textbf{\makecell{Adversary’s\\ Knowledge}}} & \multicolumn{2}{c}{\textbf{Unlearn Methods}} & \multicolumn{2}{c}{\makecell{\textbf{Unlearned Data} \\ \textbf{Modification}}} \\  
		 \cmidrule(l){2-3}   \cmidrule(l){4-5} \cmidrule(l){6-7}
		 &  \makecell{Training\\ procedure} & \makecell{White-box\\ access to model} & \makecell{Exact\\unlearning}&  \makecell{Approximate\\unlearning} & \makecell{Adding \\ perturbation} & \makecell{Remain \\ unchanged}\\
		\midrule
		\makecell{Slow-down\\ Unlearning Attacks~\cite{r14}} & \fullcirc &\fullcirc &\emptycirc &\fullcirc & \fullcirc & \fullcirc\\ \midrule
		
		\makecell{Camouflaged\\Poisoning Attacks~\cite{r28}} & \fullcirc & \fullcirc& \fullcirc& \emptycirc& \emptycirc & \emptycirc\\
	\midrule
		\makecell{Over-unlearning Attacks~\cite{hu2023duty}} & \emptycirc & \emptycirc & \emptycirc& \fullcirc&\fullcirc &\fullcirc \\
		\midrule
		
		\makecell{\textbf{Unlearning Usability}\\ \textbf{Attack (Ours)}} & \emptycirc & \emptycirc & \emptycirc&\fullcirc &\emptycirc &\emptycirc\\

		\bottomrule
	\end{tabular}
}
\label{tb1}
\end{table}


\section{Threat Model}\label{sec:threatmodel}
In this section, we describe the threat model of unlearning usability attacks, aiming to crash the unlearned model's utility in an automatic machine unlearning pipeline with hash verification. We note that the machine unlearning pipeline may also be equipped with an automatic defensive mechanism to detect poisoned data.



We anticipate the attack scenario where users contribute their data to train a machine learning model. A model developer is responsible for training the model and transferring the ownership of the model to a service provider we call the model owner. The model owner deploys the model to provide services to the users who contributed training data as well as other ``consumer'' users who want to leverage the prediction ability of the model. The attack scenario practically simulates the Machine Learning as a Service (MLaaS) environment: the model owner deploys the model in the cloud to provide services for end-users, making profits. The users who contribute the training data can submit unlearning requests to the model owner, as they have the right to delete their data. We consider that among the many ``contributor'' users, there exist one or several malicious users who intentionally wish to destroy the model by misusing their authorized rights. We detail the goals, capability, and knowledge of the malicious user, the model developer, and the service provider as follows.

\noindent \textbf{Malicious User.} \textit{i)} \textit{Goals}. A malicious user is an attacker who aims to invalidate the well-performed original model through machine unlearning. The motivation of the attacker can be complex and multifaceted, e.g., by disrupting the model owner's normal business (i.e., the model utility), the attacker may provide a competitive advantage to rival organizations. \textit{ii) Capability.} The capability of the attacker is similar to that of a normal contributor user: providing training data for training the model and requesting the model owner to unlearn exactly the data provided by the attacker. The difference between the attacker and the normal contributor user is that the training data provided by the attacker is deliberately generated to achieve the goal of the attack. \textit{iii) Knowledge}. To simulate a practical attacker, we consider the attacker has only the knowledge of the training data but without the knowledge of the target model and the unlearning method implemented by the service provider. This setting significantly distinguishes our work from existing works investigating the unlearning vulnerabilities (as detailed in Section~\ref{sec:related-work}). The knowledge of the training data can be from the attacker himself, as his data consists partially of the training dataset. This knowledge can also be from the model developer or other normal users who collude with the attacker. The worst case is that, beside his own data, the attacker has no knowledge of the training data. We will show even under the worst case, the attacker can still successfully perform the unlearning usability attacks effectively. 

\noindent \textbf{Model Developer.} The model developer belongs to an outsourced company that provides the business of training machine learning models. The model developer can access the training data and uses advanced training algorithms to train the model. After training, the ownership of the model is transferred to the service provider for making profits. Although the developer is trustful in training the model, the developer may collude for profits, e.g., selling the information of the training data for financial gain.

\noindent \textbf{Service Provider.} The service provider is responsible for maintaining the model by fulfilling the unlearning requests from contributor users. To prevent malicious unlearning requests that upload perturbed unlearned data, the model owner has previously asked the model developer to prepare a hash table of the training data. If the hash value of the uploaded unlearned data can not match the hash table, the model owner can reject the unlearning requests. This strategy invalidates previous attacks~\cite{hu2023duty,marchant2022hard} based on adding perturbations to the unlearning data.

\begin{figure}[tb]
	\centering
	\includegraphics[width=1\columnwidth]{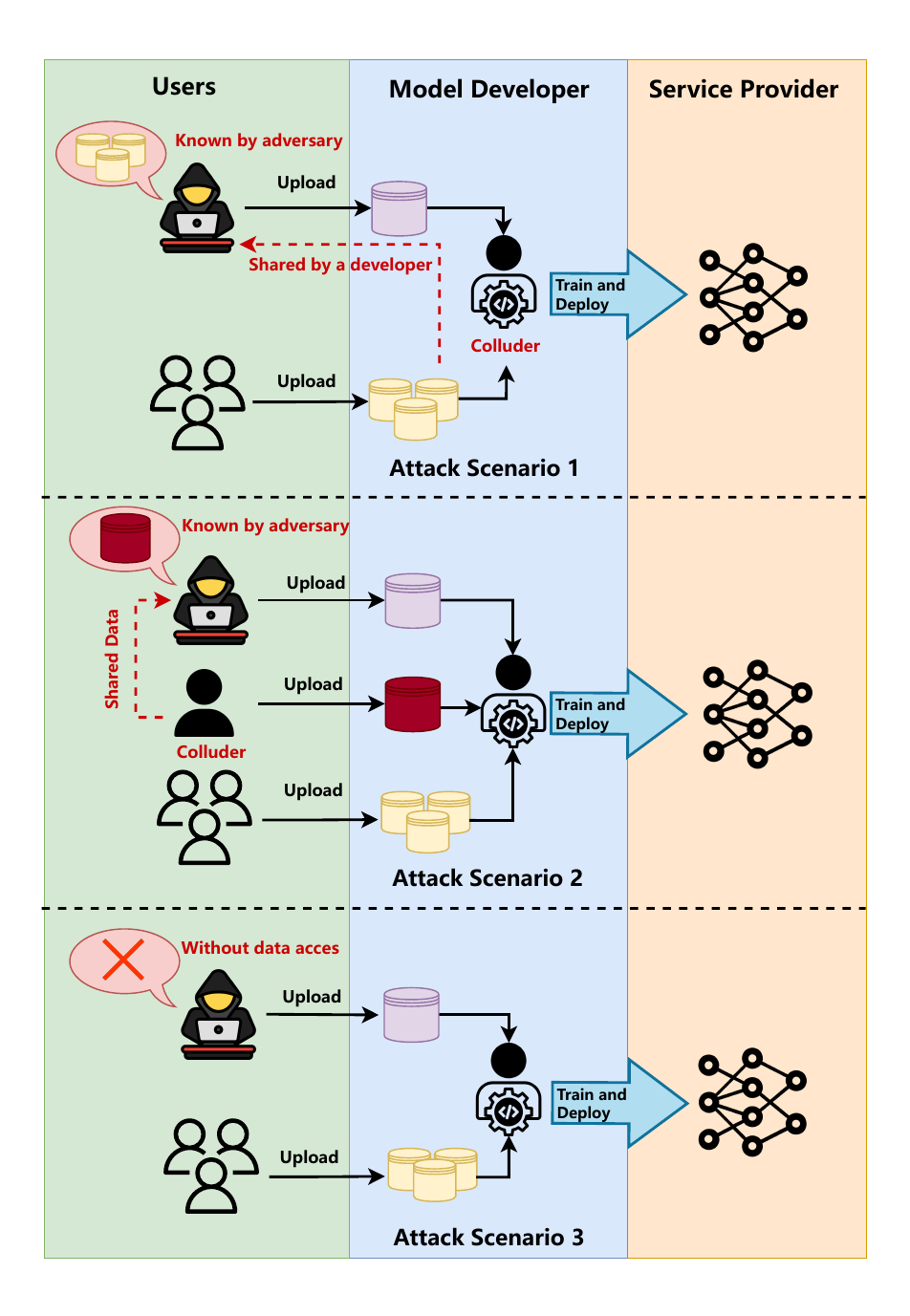}
	\caption{The three attack scenarios in unlearning usability attacks.}
	\label{diffscenario}
\end{figure}

\noindent \textbf{Detailed Attack Scenario.} We consider three attack scenarios where the attacker has different attack knowledge of the model's training data, as depicted in Figure~\ref{diffscenario} to perform unlearning usability attacks in machine unlearning. \ul{We note that Scenario 2~\&~3 may be more practical than Scenario 1, but for completeness, we consider all three scenarios.}

\noindent $\bullet$ \textbf{Scenario 1: Full Knowledge of the Training Data.} This scenario simulates a most informative attack scenario: the malicious user colluding with one model developer. Specifically, before training the model, the model developer has collected training data from different users. Thus, the model developer has the full knowledge of the training data. A malicious user can collude with the developer to obtain knowledge of the training data. Through such collision, the malicious user can gain access to the whole training dataset, thereby compromising the utility of the model through machine unlearning. Although this scenario seems too ideal for the attacker, it serves as an exploration of the upper bound of the proposed attacks.


\noindent $\bullet$ \textbf{Scenario 2: Partial Knowledge of the Training Data.} This scenario simulates the malicious user colluding with a few normal users who contribute their data to train the model. In this case, the malicious user can only have partial knowledge of the training dataset. This scenario is more practical in reality, which helps to understand how the model's utility can be damaged through machine unlearning in practice.


\noindent $\bullet$ \textbf{Scenario 3: No Knowledge of the Training Data.} This scenario simulates the most challenging setting for the malicious user: having no prior knowledge of the training data of the model. This can happen because users consider their personal data private and do not want to disclose it to others. This setting is the most strict setting for a malicious user, and it serves as an exploration of the lower bound of the proposed attack.






\section{Methodology}
In this section, we introduce our attack strategy on machine unlearning. We begin by articulating the desired impact of the attack and providing a formal definition in the Problem Statement. Subsequently, in the Problem Formulation, we describe our method for achieving the attack objective with benign data.

\noindent
\textbf{Problem Statement.} Our aim is to attack machine unlearning with the objective of inducing over-unlearning. Each data sample $(x, y)$ consists of multidimensional features $x$ and a label $y$. The neural network is denoted by the function $\psi_{\theta}(\cdot)$, which takes the sample $x$ as input and outputs the label $y$. The test set $D_{\text{test}} = {(s_1, t_1), \ldots, (s_{|N|}, t_{|N|})}$ is deployed on the server to evaluate model performance.

$\mathcal{A}(\cdot,\cdot)$ is a unlearning method, and $\psi_\theta$ represents the trained model. We denote ${D_u}$ as data uploaded by normal users, where ${D_u} \subset {D_\text{train}}$, and ${D_m}$ as data uploaded by malicious users, where ${D_m} \subset {D_\text{train}}$. 


A normal user sends an unlearning request, and the model executes \(\mathcal{A}(\psi_{\theta}, D_u)\) to perform machine unlearning. Subsequently, the model parameters are updated to \(\psi_{\theta_u}\). Similarly, a malicious user sends an unlearning request, and the model executes \(\mathcal{A}(\psi_{\theta}, D_m)\) to perform unlearning, resulting in the model parameters being updated to \(\psi_{\theta_m}\). \(\alpha_{u}\) represents the accuracy of \(\psi_{\theta_u}\) on \(D_{\text{test}}\), and \(\alpha_{m}\) represents the accuracy of \(\psi_{\theta_m}\) on \(D_{\text{test}}\). According to \cite{hu2023duty}, if the utility of ${\psi_{\theta_m}}$ on $D_{test}$ is not greater than that of ${\psi_{\theta_u}}$ on $D_{test}$, {\em i.e.,} if $\alpha_{m} < \alpha_{u}$, it is termed as a situation of over-unlearning.





\noindent
\textbf{Problem Formulation.} According to over-unlearning, the loss of model \(\psi_{\theta_m}\) on the \(D_{\text{test}}\) after over-unlearning will exceed that of model \(\psi_{\theta_u}\) after normal unlearning. Consequently, the utility of model \(\psi_{\theta_m}\) is lower than that of model \(\psi_{\theta_u}\). To achieve the goal of over-unlearning, we utilize the problem objectives outlined in \cite{r22, r23, r24, r25, r26, r27}. We represent our attack data as \(\mathcal{M}=\{(m_1,y_1),\ldots,(m_{|\mathcal{M}|},y_{|\mathcal{M}|})\}\), and the dataset known to malicious users as \(\mathcal{T}=\{(x_1, y_1),\ldots,(x_{|\mathcal{T}|}, y_{|\mathcal{T}|})\}\). We describe our attack data as 'Informative Benign Data,' where a small amount of data can demonstrate performance comparable to that of a substantial dataset. The formula is presented below:

\begin{equation}
	\label{eq2} 
	{\mathbb{E}_{x \sim {P_D}}}[\ell({\psi _{{\theta ^{\mathcal{M}}}}}(x),y)]  \simeq  {\mathbb{E}_{x \sim {P_D}}}[\ell({\psi _{{\theta ^{\mathcal{T}}}}}(x),y)]
\end{equation}

where ${P_D}$ represents the distribution of the training data, $\ell$ denotes the loss function (such as cross-entropy loss), $\psi$ is a deep neural network characterized by the parameters $\theta$, and $\theta^{\mathcal{T}}$ and $\theta^{\mathcal{M}}$ are the networks trained on datasets known to malicious users $\mathcal{T}$ and on synthetic data $\mathcal{M}$, respectively.

\noindent\textbf{Informative Benign Data.} To achieve the goal set in Formula 2 with a minimal amount of data, denoted by \(\mathcal{M}\), we aim to synthesize \(\mathcal{M}\) to closely approximate the true training data distribution. To enhance the informativeness of these samples, we employ data distribution matching methods \cite{r27}. Given the high dimensionality of training images, accurately estimating the real data distribution \(P_D\) is costly and imprecise. Therefore, we reduce each training image \(x \in \mathfrak{R}^d\) to a lower-dimensional space using parameterized functions \(\varphi_{\theta} : \mathfrak{R}^d \rightarrow \mathfrak{R}^{d'}\), where \(d' \ll d\). We then use Maximum Mean Discrepancy (MMD) \cite{r29} to estimate the distance between the actual and synthesized data distributions. As access to the ground-truth data distributions is unavailable, we rely on the empirical estimate of MMD:


\begin{equation}
	\label{eq3} 
	\mathbb{E}_{\theta \sim P_\theta} \| \frac{1}{|\mathcal{T}|} \sum_{i = 1}^{|\mathcal{T}|}\varphi_\theta(x_i) - \frac{1}{|\mathcal{M}|} \sum_{j = 1}^{|\mathcal{M}|}\varphi_\theta(m_j)  \|^{2}
\end{equation}

Where \(P_\theta\) represents the distribution of network parameters, and \(\theta\) denotes the parameters for \(\varphi_\theta\). The expectation integrates over all possible configurations of model parameters \(\theta\), thereby enhancing the robustness, generalization, and reliability of the optimization process. Following the approaches outlined in \cite{r30, r27}, we apply differentiable Siamese augmentation \(\mathcal{E}(\cdot, w)\) to both real data and \(\mathcal{M}\). During training, we randomly sample augmentations for real and synthetic minibatches, where \(w \sim \Omega\) represents augmentation parameters such as rotation angles and random cropping. By incorporating data augmentation in training deep neural networks, the learned \(\mathcal{M}\) benefits from semantic-preserving transformations and gains spatial knowledge about the samples. Finally, the optimization problem referenced in formula \ref{eq4} for \(\mathcal{M}\) is transformed to achieve the objectives specified in formula \(\ref{eq2}\).

\begin{equation}
	\label{eq4} 
	\mathop {\min }\limits_m \mathbb{E}_{\theta \sim P_\theta} \| \frac{1}{|\mathcal{T}|} \sum_{i = 1}^{|\mathcal{T}|} \varphi_\theta(\mathcal{E}(x_i,w)) - \frac{1}{|\mathcal{M}|} \sum_{j = 1}^{|\mathcal{M}|} \varphi_\theta(\mathcal{E}(m_j,w))  \|^{2}
\end{equation}

We sample \(\theta\) and learn \(\mathcal{M} = \{m_j\}_{j=1}^{|\mathcal{M}|}\) by minimizing the discrepancy between distributions in various embedding spaces. Through the optimization objective in Formula \ref{eq4}, the resulting \(\mathcal{M}\) will acquire a substantial amount of data information.

\begin{figure}[tb]
	\centering
	\includegraphics[width=1\columnwidth]{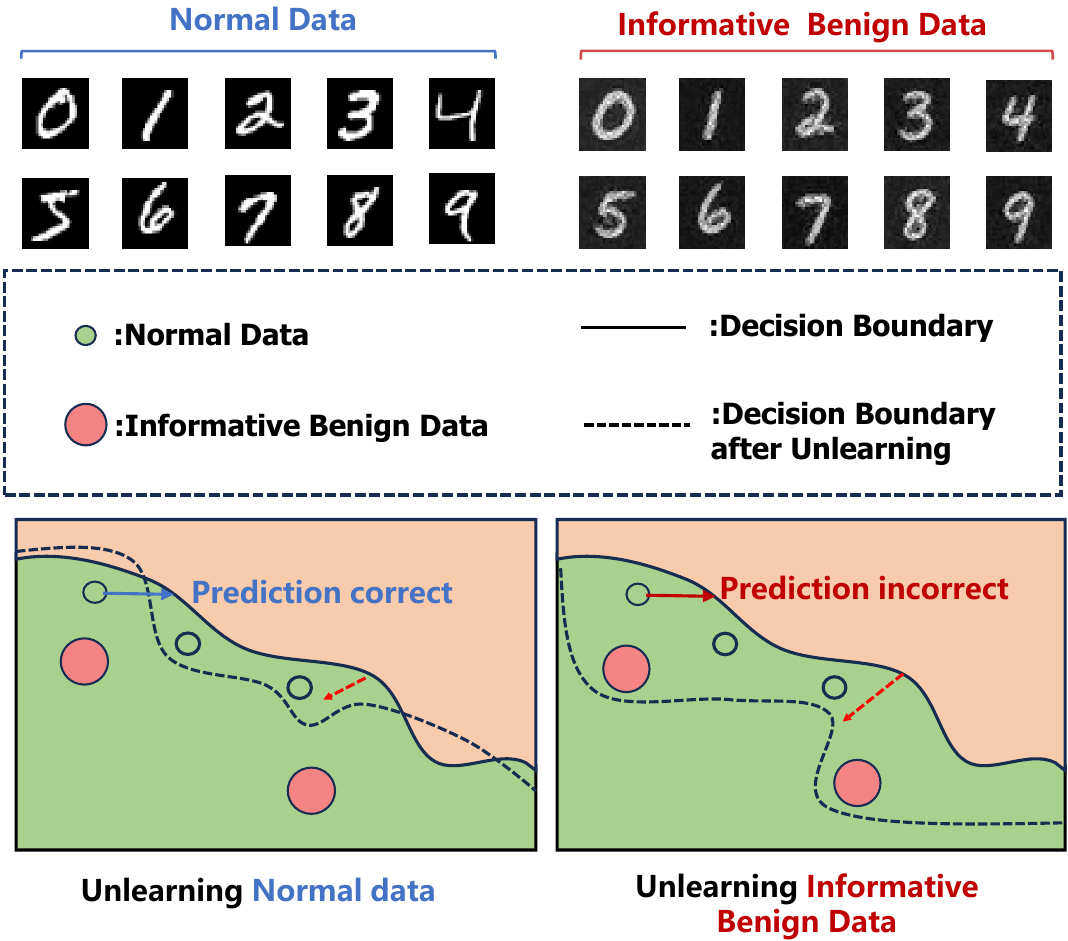}
	\caption{The distinction between normal data and informative benign data after the unlearning process.}
	\label{fig1}
\end{figure}

In Figure \ref{fig1}, we demonstrate the differences between unlearning normal data and unlearning informative benign data. Notably, unlearning the informative benign data results in a significant shift in the neural network's decision boundary, whereas unlearning normal data has a less pronounced effect. This variance stems from the richer information of the informative benign data, which is close to the centroids of the distribution.
When malicious users request unlearning of these data points, the model consequently loses crucial information that is integral to many normal samples. Table \ref{diffdata} further details the distinctions among normal data samples, informative benign data, and poisoned samples. We observe that informative benign data differs from traditional poisoned samples in that it does not harm the performance of the network. Compared to normal data samples, it is more efficient, thereby making the use of informative benign data for network training acceptable. 




\begin{table}[ht]
	\centering
	\caption{Differences between normal data, informative benign data, and poisoned data, \fullcirc: signifies complete possession of the feature, \emptycirc: indicates the absence of the feature, and \halfcirc: suggests a conditional or partial presence of the feature.}
	\resizebox{0.48\textwidth}{!}{
	\begin{tabular}{cccc}
		\hline
  
		 &Clean Label& \makecell{Harm Network \\Accuracy} & \makecell{Highly effective}\\ \hline \\ [-2ex]

   Normal Data& \fullcirc & \emptycirc  &\emptycirc \\

   Informative Benign Data& \fullcirc & \emptycirc & \fullcirc\\
   
   Poisoned Data& \halfcirc & \fullcirc & \emptycirc \\

		\hline
	\end{tabular}
 }
\label{diffdata}
\end{table}

\section{Experimental Settings}\label{sec:experiment_settings}

\noindent
\textbf{Datasets.}  We have considered three attack scenarios, where the attacker has varying levels of knowledge about the training data. In scenario 1 and 2, we assess the impact of our attack method across various unlearning techniques on four datasets, {\em i.e.,} MNIST \cite{r31}, Fashion MNIST (FMNIST), CIFAR10, and CIFAR100 \cite{r32}. Both MNIST and FMNIST consist of 60,000 grayscale training images, each sized 28x28 and divided into 10 classes. CIFAR10 and CIFAR100 comprise 50,000 32×32 training images representing 10 and 100 object categories, respectively. In scenario 3, we assess the impact of our attack method across various unlearning techniques on MNIST-M and Tiny-ImageNet datasets. MNIST-M \cite{ganin2015unsupervised}, a variant of the MNIST dataset, combines digit images with natural scenes to increase diversity and simulate real-world complexity. This enhanced version is used to assess model robustness and generalization by introducing challenging backgrounds and noise that differ from the original MNIST dataset. MNIST-M contains 60,000 colorful training images sized 28×28. Tiny-ImageNet, a subset of the ImageNet dataset, is designed for image classification and includes images from 200 categories.

\smallskip
\noindent \textbf{Informative Benign Data Configuration.} We have specifically conducted the following experimental setups for each of these scenarios.

\noindent $\bullet$ \textbf{Scenario 1: User-Developer Collusion.} Following our objective formula \ref{eq4} and adopting the approach from \cite{r27}, we generate informative benign data using the same ConvNet architecture as \cite{r23}. For each category in the MNIST \cite{r31}, FMNIST, and CIFAR10 datasets, we generate 10 informative images. In the case of CIFAR100 \cite{r32}, we produce 1 informative image for each category, serving as informative benign images.

\noindent $\bullet$ \textbf{Scenario 2: User Collusion.} We introduce a metric called "Dataset Knowledge" (Dst. Kwl.), which is defined as the proportion of the whole dataset known by the adversary to quantify the adversary's knowledge. In the MNIST and FMNIST datasets, we set the minimum Dst. Kwl. to 1\%, while for the CIFAR10 and CIFAR100 datasets, it is set to 2\%. We then increased the Dst. Kwl. to 5\% and 10\% for these four datasets, respectively. Subsequently, adversaries generate informative benign data based on the specified Dst. Kwl. For each setting, we produce 10 informative images for each category in the MNIST, FMNIST, and CIFAR10 datasets. For the CIFAR100 dataset, however, we generate one informative image per category.

\noindent $\bullet$ \textbf{Scenario 3: No Collusion.} In Scenario 3, because we don't have access to other users' data, we use an out-of-distribution (POOD) dataset to create helpful informative benign data. POOD data are examples that the model hasn't seen during training but might encounter in real life. For instance, while a model trained on MNIST might perform well with that dataset, it could struggle with MNIST-M because of different image distributions. Importantly, the POOD data we use here are different from what the users trained on. For the MNIST-M attack, we used MNIST as our POOD data. When attacking Tiny-ImageNet, we picked classes like cats, dogs, and cars that overlap with CIFAR10 and used CIFAR10 as the POOD dataset for our attacks. We generated 20 informative benign images from MNIST for each category in MNIST-M and 5 informative benign images from CIFAR10 for each category in Tiny-ImageNet.

\noindent
\textbf{Unlearning Benchmarks.} We assess the effectiveness of our proposed methods for over-unlearning across four benchmark unlearning techniques. 

\noindent $\bullet$ \textbf{The first-order based unlearning method \cite{r12}}. This method derives the gradient updates using a first-order Taylor series expansion of the model \(\psi_{\theta}\). The model update target formula is shown as \ref{eqcite12}. Here, $\tau$ is a pre-defined unlearning rate. The dataset \(D\), consisting of elements \((x, y)\) targeted for unlearning. The corresponding unlearned \(\widetilde D\) dataset defined by \(\widetilde{x} = (x + \delta_x, y)\), where \(\delta_x\) is the applied perturbation, and \(\psi_{\theta^{*}}\) is the unlearned model. $\ell$ is a loss function (e.g., cross-entropy)

\begin{equation}
	\label{eqcite12} 
	\psi_{\theta^{*}} \leftarrow  \psi_{\theta} - \tau(\sum\limits_{\widetilde x \in \widetilde D} {{\nabla _\theta }\ell(\psi_{\theta}(\widetilde{x}_i), y) - \sum\limits_{x \in D} {{\nabla _\theta }\ell(\psi_{\theta}({x}_i), y )} } )
\end{equation}

\noindent $\bullet$ \textbf{The second-order based unlearning method \cite{r12}}. This method utilizes the inverse Hessian matrix, derived from the second-order partial derivatives, to modify the parameters of the original model, thus creating the unlearned model. The unlearned model can be described as follows in equation \ref{eqsecond}: \(H_{{\psi_\theta}}^{-1}\) is the inverse Hessian matrix, \(\ell\) is a loss function, and \(\psi_{\theta^*}\) is the unlearned model.

\begin{equation}
	\label{eqsecond} 
	\psi_{\theta^{*}} \leftarrow  \psi_{\theta} - H_{{\psi _\theta }}^{ - 1}(\sum\limits_{\widetilde x \in \widetilde D} {{\nabla _\theta }\ell(\psi_{\theta}(\widetilde{x}_i), y) - \sum\limits_{x \in D} {{\nabla _\theta }\ell(\psi_{\theta}({x}_i), y )} } )
\end{equation}


\noindent $\bullet$ \textbf{The negative gradient unlearning method \cite{r8}}. This method unlearns the samples by simply maximizing their loss. The model update target formula is shown in \ref{maxloss}. In this context, $\tau$ represents the unlearning rate, and $D$ is the dataset targeted for unlearning. $\ell$ denotes a loss function, and $\psi_{\theta^*}$ is the unlearned model.

\begin{equation}
	\label{maxloss} 
	\psi_{\theta^{*}} \leftarrow  \psi_{\theta}  + \tau \sum\limits_{x \in D} {{\nabla _\theta }\ell(\psi_{\theta}({x}_i), y )} )
\end{equation}

\noindent $\bullet$ \textbf{The amnesiac unlearning method \cite{r18}}. This method uses a fine-tuning approach carefully designed to manage both the training and unlearning stages of the model. This allows the model to effectively "unlearn" previously absorbed images through fine-tuning. As outlined in formula \ref{eq6}, $\psi_{\theta_{initial}}$ represents the untrained model, $\Delta \psi_{\theta_{e,b}}$ denotes the parameter updates generated by the entire training dataset, where $e$ denotes total epochs and $b$ denotes the number of batches in each epoch, and $\Delta \psi_{\theta_{u,b}}$ signifies the updates produced by data designated for unlearning during training, with $ub \in UB$ representing the batch produced by the data for unlearning.

\begin{equation}
	\label{eq6} 
	\mathcal{A}({\psi _\theta }, D) = {\psi _{{\theta _{initial}}}} + \sum\limits_{e = 1}^E {\sum\limits_{b=1}^B {\Delta {\psi _{{\theta _{e,b}}}} - } } \sum\limits_{ub = 1}^{UB} {\Delta {\psi _{{\theta _{u,b}}}}}
\end{equation}

In each of the mentioned unlearning methods, the MLaaS Server cannot use additional user data to fine-tune the model. This is crucial for practical applications, as reusing user data for fine-tuning each time an unlearning method is invoked raises privacy concerns and increases the risk of DDoS attacks. We unlearn both normal data and our Informative Benign Data for the same number of epochs. To ensure optimal unlearning effects, we initially fine-tune the unlearning methods on normal data, achieving the most thorough unlearning without significantly impacting accuracy. Subsequently, we use these parameters to unlearn Informative Benign Data. We employed ResNet18 \cite{r33} as the foundational network architecture for our study.


\noindent
\textbf{Defense Configuration.} To evaluate the resistance of our informative benign samples against defenses, we compared them to classic dirty-label (BadNets \cite{rbad}, Blend \cite{rblend}) and clean-label (LC \cite{rlc}) backdoor attack samples, illustrating the differences in defense resistance. Defense methods are broadly classified into active and passive types \cite{r40}. For passive defense, we tested our informative benign samples' resistance against Spectral Signature \cite{r39} with 500 attack samples and further assessed resistance against SPECTRE \cite{r37} using fewer than 250 samples. Additionally, we examined their resistance to Strip \cite{r41} and TaCT \cite{r38} defenses. Lastly, we tested against the active defense CT \cite{r42}. The aforementioned defense methods all utilize the True Positive Rate (TPR) as the evaluation metric, calculated as $\frac{{{D_{detect}} \cap {D_{poison}}}}{{|{D_{poison}}|}}$. Our approach follows the same settings as outlined in the original paper for the aforementioned methods.


\noindent
\textbf{Metric.} The primary evaluation metrics are as follows:

\noindent\textit{i) Test accuracy:} This metric aims to measure the model's ability to deliver accurate predictions, given the limited server capacity to evaluate unlearned models. It serves as the most practical indicator, determined using a test dataset.

\noindent\textit{ii) Train accuracy:} This metric aims to assess this metric, gauging the impact of Informative Benign Data on the training set, which potentially contains critical dataset information. The loss of such data is expected to lower the training set's accuracy.

\noindent\textit{iii) Budget:} Budget aims to assess the model's difficulty in unlearning  data points. According to \cite{r12} and formula \ref{eqcite12}, optimal unlearning can be achieved by identifying the data points \(x\) that need to be unlearned and applying the optimal perturbation \(\delta_x\). Consequently, we calculate the perturbation \(\delta_x\) for each \(x\) to realize the optimal unlearning effect.  We denote the sum of perturbations (budget) required for each data point \(x\) as \(\sum\limits_i {|{\delta _{x_i}}|} \).

\noindent\textit{iv) Parameter Update Metric:} This metric aims to evaluate the distinct contributions of each data type to the model. In Amnesiac Unlearning, as outlined in formula \ref{eq6}. In this metric, we independently recorded the parameter updates generated by normal and informative benign data to ensure no overlap between them. Subsequently, during the unlearning process, we removed the updates generated by normal data and informative benign data, respectively.

\section{Attack Performance} 
We follow the experimental settings in Section~\ref{sec:experiment_settings} to evaluate our attack and provide the results in this section. All the evaluation results for different machine unlearning techniques indicate unlearning informative benign data will be much more harmful than unlearning normal data. 

\textit{Attack Performance in Scenario 1.}
In the scenario of user-developer collusion (refer to the settings in Section~\ref{sec:experiment_settings}), 

the model performance across different datasets before unlearning is shown in Table~\ref{tbmetric}. The changes in model accuracy after unlearning normal or informative data are shown in Table~\ref{tb2}. We observe that unlearning informative data can cause a decrease in model accuracy of at least 20\% by utilizing only 0.2\% of the data.

For amnesiac unlearning, when the model is instructed to erase updates generated by informative data, the model accuracy significantly drops to 44.33\%, 10.25\% and 3.17\% on MNIST, CIFAR-10, and CIFAR-100, respectively. In comparison, unlearning normal data only drops the accuracy to 99.04\%, 86.01\% and 47.38\% on those datasets.
When unlearning normal data, first-order and second-order unlearning methods can erase the influence of data features with negligible impact on model accuracy. However, when unlearning informative data, we observed a decrease in accuracy of about 30\% on MNIST and more than 20\% on CIFAR-10 and CIFAR-100 datasets.
For the negative gradient method, we also observed that unlearning normal data does not significantly decrease the model’s accuracy. However, for our informative benign data, both test and training accuracy dropped by at least 20\% across all datasets. We believe that when performing gradient ascent on informative benign data, the updates affect more crucial parameters in the model, leading to a decrease in model performance.

\begin{table}[tb]
	\centering
	\caption{In scenario 1, model's performance across different datasets before unlearning.}
	\resizebox{0.4\textwidth}{!}{
		\begin{tabular}{ccc}
			\hline
			\makecell{Metric $\rightarrow$  \\ Dataset $\downarrow$}  &\makecell{Acc. on $D_{test}$} & \makecell{Acc. on $D_{train}$}\\ \hline
			
			\makecell{MNIST \cite{r31} }  & 99.06 &99.44\\  
			\makecell{FMNIST} & 93.56& 96.71\\
	
			\makecell{CIFAR10} &89.52 &97.13\\
			
			\makecell{CIFAR100 \cite{r32}}&64.38 &88.09 \\ \hline

		\end{tabular}
	}
	\label{tbmetric}
\end{table}



			

\begin{table*}[tb]
	\centering
	\caption{In scenario 1, User-Developer Collusion, the comparison involves the model's performance in $D_{test}$ accuracy and $D_{train}$ accuracy on the MNIST, CIFAR10, and CIFAR100 datasets when unlearning normal data and informative benign data, respectively.}
	\resizebox{1\textwidth}{!}{
		\begin{tabular}{cccccccccc}
			\hline
			
			  \multicolumn{1}{c}{}&  \multicolumn{4}{c}{ \raisebox{-1ex}{ Unlearning Normal Data}}  & \multicolumn{4}{c}{\raisebox{-1ex}{Unlearning Informative Benign Data}} \\[2ex] \cline{2-5} \cline{7-10} \\ [-2ex]
			
			  \multicolumn{1}{c}{Metrics} & \makecell{\textbf{Amn. Unl. \cite{r10}} } & \makecell{\textbf{Neg. Grad.}} & \makecell{\textbf{First-Order \cite{r12}} } & \makecell{\textbf{Second-Order \cite{r12}} } & &
			  
			   \makecell{\textbf{Amn. Unl. \cite{r10}}} & \makecell{ \textbf{Neg. Grad.}} &\makecell{\textbf{First-Order \cite{r12}}} & \makecell{\textbf{Second-Order \cite{r12}}} \\  \hline 
			  
			  \multicolumn{10}{c}{\textbf{\raisebox{-1ex} {MNIST \cite{r31} results, 0.17\% percentage (10 images per class)}}} \\ [2ex] \hline
			 
			  \makecell{Acc. on $D_{test}$} & 99.04 & 99.05 & 99.02 & 99.07 & & 44.33 (54.71$\downarrow$) & 76.1 (22.95$\downarrow$) & 60.18 (38.84$\downarrow$) & 51.73 (47.34$\downarrow$) \\
			  \makecell{Acc. on $D_{train}$} & 99.07 & 99.41  &  99.32 & 99.44 & &   42.29 (56.78$\downarrow$)& 76.67 (22.74$\downarrow$) & 59.52 (39.8$\downarrow$) & 51.52 (47.92$\downarrow$)\\ \hline

            \multicolumn{10}{c}{\textbf{\raisebox{-1ex} {FMNIST \cite{r31} results, 0.17\% percentage (10 images per class)}}} \\ [2ex] \hline
			 
			  \makecell{Acc. on $D_{test}$} & 89.51 & 93.99 & 93.48 & 93.60 & & 43.47 (46.04$\downarrow$) & 49.53 (44.46$\downarrow$) & 64.88 (28.6$\downarrow$) & 62.73 (30.87$\downarrow$) \\
			  \makecell{Acc. on $D_{train}$} & 92.38 & 97.14  &  96.83 & 96.68 & &  44.53 (47.85$\downarrow$)& 50.62 (46.52$\downarrow$) & 67.03 (29.8$\downarrow$) & 66.91(29.77$\downarrow$)\\ \hline
			  
			  \multicolumn{10}{c}{\textbf{\raisebox{-1ex} {CIFAR10 \cite{r32} results, 0.2\% percentage (10 images per class)}}} \\ [2ex] \hline
			  
			  \makecell{Acc. on $D_{test}$} & 86.01 & 89.39 &89.30  &89.58  &  & 10.25 (75.76$\downarrow$) & 69.79 (19.6$\downarrow$)& 67.36 (21.94$\downarrow$) & 57.51 (32.07$\downarrow$) \\
			  
			  \makecell{Acc. on $D_{train}$} & 92.81 & 96.89  & 96.99 &97.12 &  &   10.27 (82.54$\downarrow$) & 75.64 (21.25$\downarrow$) & 73.31 (23.68$\downarrow$) & 61.19 (35.93$\downarrow$)\\ \hline
			  
			  
			  \multicolumn{10}{c}{\textbf{\raisebox{-1ex} {CIFAR100 \cite{r32} results, 0.2\% percentage (1 images per class)}}} \\ [2ex] \hline
			  
			  \makecell{Acc. on $D_{test}$} & 47.38 & 63.41 & 63.01 & 64.32 & & 3.17 (44.21$\downarrow$) & 41.76 (21.65$\downarrow$) & 35.62 (27.39$\downarrow$) & 42.86 (21.46$\downarrow$) \\
     
			  \makecell{Acc. on $D_{train}$} & 58.75 & 86.63  &  86.17 & 87.94 & & 3.18 (55.57$\downarrow$) & 54.86 (31.77$\downarrow$) & 46.17 (40.0$\downarrow$) & 55.75 (32.19$\downarrow$)\\ 
			
			\hline
		\end{tabular}
	}
	\label{tb2}
\end{table*}

To further mitigate privacy risks, model owners might execute multiple unlearning rounds to ensure complete data erasure and protect user privacy.

\begin{figure}[tb]
	\centering
	\includegraphics[width=1\columnwidth]{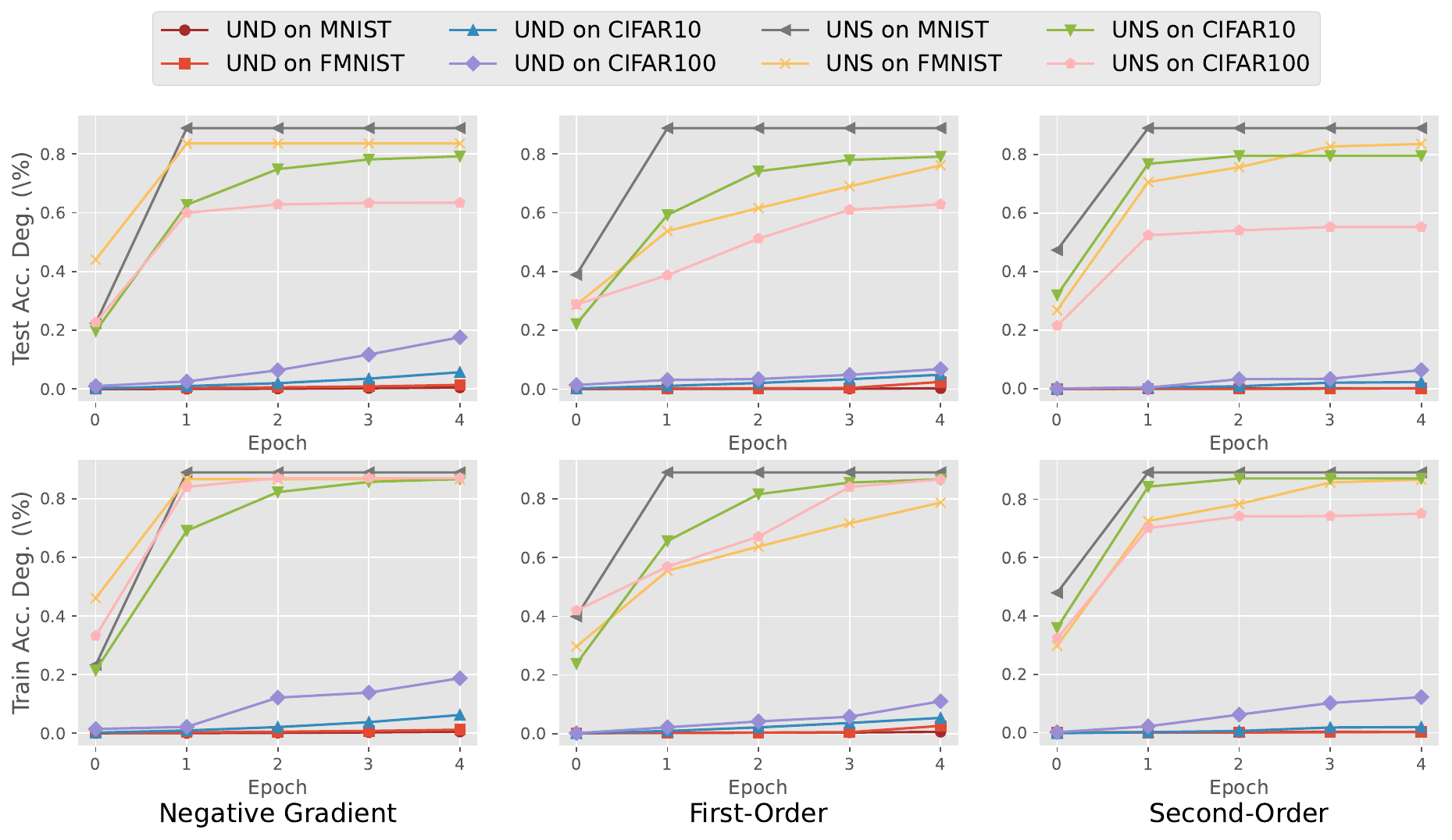}
	\caption{Model performance comparison between normal and informative benign data under multiple rounds of unlearning.}
	\label{figacc_de}
\end{figure}

Our evaluation indicates that, while multi-round fine-tuning for unlearning normal data preserves network stability typically results in less than a 10\% accuracy drop on all the datasets. Multi-round fine-tuning for unlearning informative datasets will lead to a significant decline in model utility, with accuracy reductions exceeding 40\%. The results of fine-tuning on both data types across MNIST, FMNIST, CIFAR-10, and CIFAR-100 are illustrated in Figure \ref{figacc_de}, where 'UND' and 'UNS' represent the unlearning of normal and informative data, respectively.

\textit{Attack Performance in Scenario 2.} For the user collision scenario, we provide the main results in Table \ref{tbscenario2per}. Specifically, the  "Before Unlearning" column shows the accuracy of the model trained on normal data and informative data generated at different levels of Dst. Kwl., which verifies that the informative data will not negatively affect model accuracy in all the cases. 

However, removing informative data led to a minimum 10\% decrease in accuracy, regardless of the adversary's knowledge. It's important to note that higher Dst. Kwl. values usually resulted in greater accuracy reductions, and when Dst. Kwl. reaches $10\%$, the attack can achieve a similar effect as attack scenario 1.

\begin{table*}[tb]
	\centering
	\caption{In scenario 2: User Collusion, we measure the impact of the attack by analyzing 'Dst. Kwl.,' which represents the portion of the dataset accessible during User Collusion. The \colorbox{gray!30}{gray} indicates a decrease in accuracy similar to that observed in scenario 1.}
	\resizebox{1\textwidth}{!}{
		\begin{tabular}{ccc|cc|cc|cc|cc}
			\hline \\[-2ex]
			&& \multicolumn{1}{c|}{ \raisebox{-1ex}{\textbf{Before Unlearning} }} & \multicolumn{2}{c|}{ \raisebox{-1ex}{\textbf{Amnesiac Unlearning \cite{r10}}}} & \multicolumn{2}{c|}{\raisebox{-1ex}{\textbf{Negative Gradient}}}  &\multicolumn{2}{c|}{\raisebox{-1ex}{\textbf{First-Order \cite{r12}}}}&\multicolumn{2}{c}{\raisebox{-1ex}{\textbf{Second-Order \cite{r12}}}}  \\ [2ex]
			
			& Dst. Kwl.(\%)& Test Acc. & Normal Acc.& Informative Acc.& Normal Acc. &  Informative Acc. & Normal Acc. &  Informative Acc. & Normal Acc. &  Informative Acc. \\ \\[-1.5ex] \hline \\ [-1.5ex]

			\multirow{4}{*}{\makecell{MNIST\\(0.17\% Percentage)}} & \multicolumn{1}{|c|}{1}&99.25 & 99.01 & 68.55 (30.46$\downarrow$)& 99.30 & 84.83 (14.47$\downarrow$) &99.2 & 85.27 (13.93$\downarrow$)& 98.99 & 84.10 (14.89$\downarrow$)\\ 
			
			& \multicolumn{1}{|c|}{5} &99.24 & 98.56 & 53.18 (45.38$\downarrow$)& 99.24 & 83.13 (16.11$\downarrow$)& 99.16 & 76.59 (22.57$\downarrow$)&99.26 & 68.79 (30.47$\downarrow$)\\

			& \multicolumn{1}{|c|}{10} &99.32 &99.10&\cellcolor{gray!30} 46.93 (52.17$\downarrow$)& 99.26 &\cellcolor{gray!30} 78.98 (20.28$\downarrow$)& 98.93 &\cellcolor{gray!30} 74.32 (24.61$\downarrow$)& 99.06 &\cellcolor{gray!30} 55.21 (43.85$\downarrow$)\\ \\[-1.5ex] \hline \\ [-1.5ex]

			\multirow{4}{*}{\makecell{FMNIST\\(0.17\% Percentage)}} &  \multicolumn{1}{|c|}{1}& 93.56& 87.19& 65.57 (21.62$\downarrow$)& 93.17 & 81.44 (11.73$\downarrow$)& 93.23& 78.49 (14.74$\downarrow$)& 93.31 & 71.12 (22.19$\downarrow$)\\ 
			
			& \multicolumn{1}{|c|}{5} &93.94& 89.87 & 46.33 (43.54$\downarrow$)& 93.90  &63.31 (30.59$\downarrow$)& 93.32 & 80.54 (12.78$\downarrow$)& 92.95 & 78.84 (14.11$\downarrow$)\\ 
			
			& \multicolumn{1}{|c|}{10} &94.46& 87.22& \cellcolor{gray!30} 44.02  (43.2$\downarrow$)& 93.53 &\cellcolor{gray!30}  54.59 (38.94$\downarrow$)& 93.44 & \cellcolor{gray!30} 77.39 (16.05$\downarrow$)& 94.01 &\cellcolor{gray!30} 65.99 (28.02$\downarrow$)\\  \\[-1.5ex] \hline \\ [-1.5ex]
			
			\multirow{4}{*}{\makecell{CIFAR10\\(0.2\% Percentage)}} & \multicolumn{1}{|c|}{2} &89.91&86.86&10.23 (76.63$\downarrow$)& 89.26  & 77.83 (11.43$\downarrow$)& 88.97& 74.84 (14.13$\downarrow$)& 89.95 & 77.63 (12.32$\downarrow$)\\ 
			
			& \multicolumn{1}{|c|}{5} & 90.2 & 84.35 &10.23 (74.12$\downarrow$)& 88.95 & 74.91 (14.04$\downarrow$)& 89.83 & 70.40 (19.43$\downarrow$)& 88.67& 71.67 (17.0$\downarrow$)\\ 
			
			& \multicolumn{1}{|c|}{10} & 89.52 & 86.01 & \cellcolor{gray!30} 10.04 (75.97$\downarrow$)& 88.68& \cellcolor{gray!30} 70.79 (17.89$\downarrow$)& 89.17 &\cellcolor{gray!30}  68.54 (20.63$\downarrow$)& 89.50 & \cellcolor{gray!30} 64.33 (25.17$\downarrow$)\\  \\[-1.5ex] \hline \\ [-1.5ex]

			\multirow{4}{*}{\makecell{CIFAR100\\(0.2\% percentage)}} & \multicolumn{1}{|c|}{2} &64.00& 43.53& 2.2 (41.33$\downarrow$)& 64.10 & 47.65 (16.45$\downarrow$)& 63.82 & 52.70 (11.12$\downarrow$)& 63.20  &  49.32 (13.88$\downarrow$)\\ 
			
			& \multicolumn{1}{|c|}{5} & 65.54 & 47.38 & 2.18 (45.20$\downarrow$)&  64.59 & 46.20 (18.39$\downarrow$)& 64.12 & 49.00 (15.12$\downarrow$)& 63.05 & 47.90 (15.15$\downarrow$)\\ 
			 
			& \multicolumn{1}{|c|}{10} & 64.49 & 46.20 & \cellcolor{gray!30} 2.18 (44.02$\downarrow$)& 63.51 & 43.55 \cellcolor{gray!30} (19.96$\downarrow$)& 64.00 & 40.31 \cellcolor{gray!30} (23.69$\downarrow$)& 63.58 & \cellcolor{gray!30} 45.28 (18.30$\downarrow$)\\ \\[-1.5ex]\hline \\ [-1.5ex]
		\end{tabular}
	}
	\label{tbscenario2per}
\end{table*}

\textit{Attack Performance in Scenario 3.} In this scenario, due to the lack of access to other users' data, we utilize POOD data (refer to Section~\ref{sec:experiment_settings}) for generating informative benign data. Table \ref{attscen3} illustrates the specific attack results, revealing that even without accessing information from other user datasets, the informative data generated by the POOD dataset still leads to a more significant decrease in accuracy during unlearning compared with normal data. For Amnesiac Unlearning, with injection rates of just 0.33\% and 1\%, unlearning the informative data generated by the POOD dataset reduces model accuracy to 82.52\% and 47.67\% on MNIST-M and Tiny-ImageNet, respectively. In comparison unlearning normal data results in model accuracy of 94.14\% and 69.67\%. For Negative Gradient, First-Order, and Second-Order unlearning methods, unlearning informative data drops the accuracy by up to about 40\% more than unlearning normal data. 

\textit{In Appendix \ref{app:abstud}, we conducted ablation studies and assessed the applicability of our approach across different architectures.}

\begin{table*}[tb]
	\centering
	\caption{In scenario 3: No Collusion, we evaluate the attack's performance by launching attacks using the POOD dataset to generate informative benign data.}
	\resizebox{1\textwidth}{!}{
		\begin{tabular}{ccccccccccc}
		\hline
			&& & \multicolumn{2}{c}{ \raisebox{-1ex}{\textbf{Amnesiac Unlearning \cite{r10}}}} & \multicolumn{2}{c}{\raisebox{-1ex}{\textbf{Negative Gradient}}}  &\multicolumn{2}{c}{\raisebox{-1ex}{\textbf{First-Order \cite{r12}}}}&\multicolumn{2}{c}{\raisebox{-1ex}{\textbf{Second-Order \cite{r12}}}}   \\  \\ [-2ex] \cline{4-11}
			
			Dataset& \makecell{Percentage(\%)}&\makecell{Before\\Unlearning}&Nor. Data& Poi. Data&Nor. Data& Poi. Data&Nor. Data& Poi. Data&Nor. Data& Poi. Data \\ \hline \\ [-2ex]  
			
			\multirow{4}{*}{MNIST-M} & 0.33& 98.57 &94.14& 82.52 (11.62$\downarrow$)& 98.54& 74.86 (23.68$\downarrow$)& 98.43&73.54 (24.89$\downarrow$)&98.19& 69.3 (28.89$\downarrow$)\\   
   
			& 0.50 &  98.43& 86.33 &47.18 (39.15$\downarrow$)& 98.42  &60.66 (37.76$\downarrow$)& 98.10& 68.22 (29.88$\downarrow$)& 98.23& 64.90 (33.33$\downarrow$)\\ 

			& 0.66 &  98.64& 83.59& 46.27 (37.32$\downarrow$)& 98.53 &51.56 (46.97$\downarrow$)&98.14 &61.48 (36.66$\downarrow$)& 98.10 &55.87 (42.23$\downarrow$)\\

			\multirow{2}{*}{Tiny-ImageNet} & 1  & 77.67& 69.67&47.67 (22$\downarrow$)& 74.33& 68.33 (6$\downarrow$)& 74.33&69.00 (5.33$\downarrow$)& 74.00& 63.33 (10.67$\downarrow$)\\   
   
			& 2  &74.33& 55.33 &39.33 (16$\downarrow$)&73.00&65.67 (7.33$\downarrow$)&72.00&65.00 (7$\downarrow$)&73.00& 65.67 (7.33$\downarrow$)\\ \\ [-2ex] \hline
		\end{tabular}
	}
	\label{attscen3}
\end{table*}

\textit{Model Retraining Without Informative Data.} We trained a model on an original dataset without injecting informative benign data and observed changes in the model's accuracy by unlearning normal data. The main results are shown in Table \ref{tbwithoutinfor}. We observed that the model's accuracy after training on the original dataset is comparable to the accuracy achieved when training on a dataset with injected informative benign data. Specific references to the accuracy after training in attack scenarios 1, 2, and 3 can be found in Table \ref{tbmetric}, Table \ref{tbscenario2per}, and Table \ref{attscen3}, respectively. This further indicates that informative benign data does not have a negative impact on the model. In these tables, we can also observe that the decrease in model accuracy when unlearning normal samples does not exceed the decrease observed when unlearning informative benign data in attack scenarios 1, 2, and 3. This further indicates that unlearning informative benign data may be more harmful to the model.

\begin{table*}[tb]
	\centering
	\caption{Train the model on a normal dataset without injecting informative benign data, and observe the change in model accuracy after unlearning the normal data.}
	\resizebox{1\textwidth}{!}{
		\begin{tabular}{cccccccccccc}
			\hline \\[-2ex]
			& &\multicolumn{2}{c}{ Before Unlearning } & \multicolumn{2}{c}{ Amnesiac Unlearning \cite{r10}} & \multicolumn{2}{c}{Negative Gradient}  &\multicolumn{2}{c}{First-Order \cite{r12}}&\multicolumn{2}{c}{Second-Order \cite{r12}}  \\  \\[-2ex]   \cline{3-12}  \\[-2ex]
			
			Dataset &\makecell{Percentage(\%)}& Train Acc.&Test Acc. &Train Acc.& Test Acc.&Train Acc.& Test Acc. &Train Acc.& Test Acc. &Train Acc.& Test Acc. \\ \\[-2ex] \hline \\[-2ex]

			MNIST & 0.17\%& 99.04 &98.69& 98.37 & 98.01 &99.03 & 98.01 &98.92 &98.70 &99.01 &98.14 \\   
		 \\[-2ex] 
			\makecell{FMNIST}  & 0.17\%&96.95& 93.59 &91.60 &88.05 &96.90 &93.93&   95.87 &93.18 & 96.22 & 93.42 \\ 
			  \\[-2ex] 
			
			CIFAR10  & 0.2\%&97.08& 89.72& 89.78 & 84.47 & 96.53 & 89.05 & 97.01 &89.60&96.93&89.55 \\ 
			 \\[-2ex]

			CIFAR100  & 0.2\% &88.13 & 64.11 & 60.16& 49.90& 85.51 & 62.87 &87.73 & 64.10& 86.59 & 63.91 \\ 
			 \\[-2ex] 
			 
			 MNIST-M  & 0.33\% &99.51 & 98.61 & 95.20 & 94.14 & 99.10 & 98.20 & 99.45 & 98.06 & 99.17 & 98.22 \\ 
			 \\[-2ex] 
			 
			 Tiny-ImageNet  & 1\% &97.33 & 74 & 90.80 & 69.67 & 95.73 & 73.33 & 97 & 74 & 95.33 & 73.33 \\ \\[-2ex]\hline
			 
		\end{tabular}
	}
	\label{tbwithoutinfor}
\end{table*}
 
\section{Analysis of Attack Effectiveness}
\begin{figure}[tb]
	\centering
	\includegraphics[width=1\columnwidth]{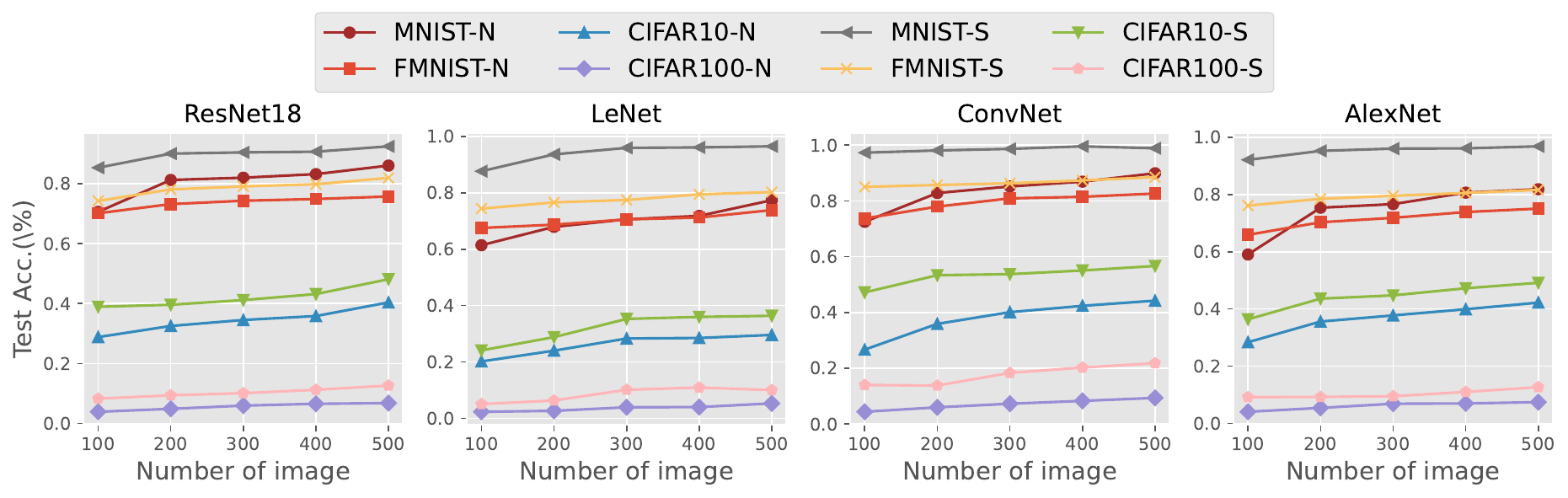}
	\caption{In scenario 1, we assess the information of informative ('S') and normal ('N') data across different networks. We train both types of data on two separate networks and then compare their accuracies to determine their relative information. }
	\label{scenario1info}
\end{figure}

\begin{figure}[tb]
	\centering
	\includegraphics[width=1\columnwidth]{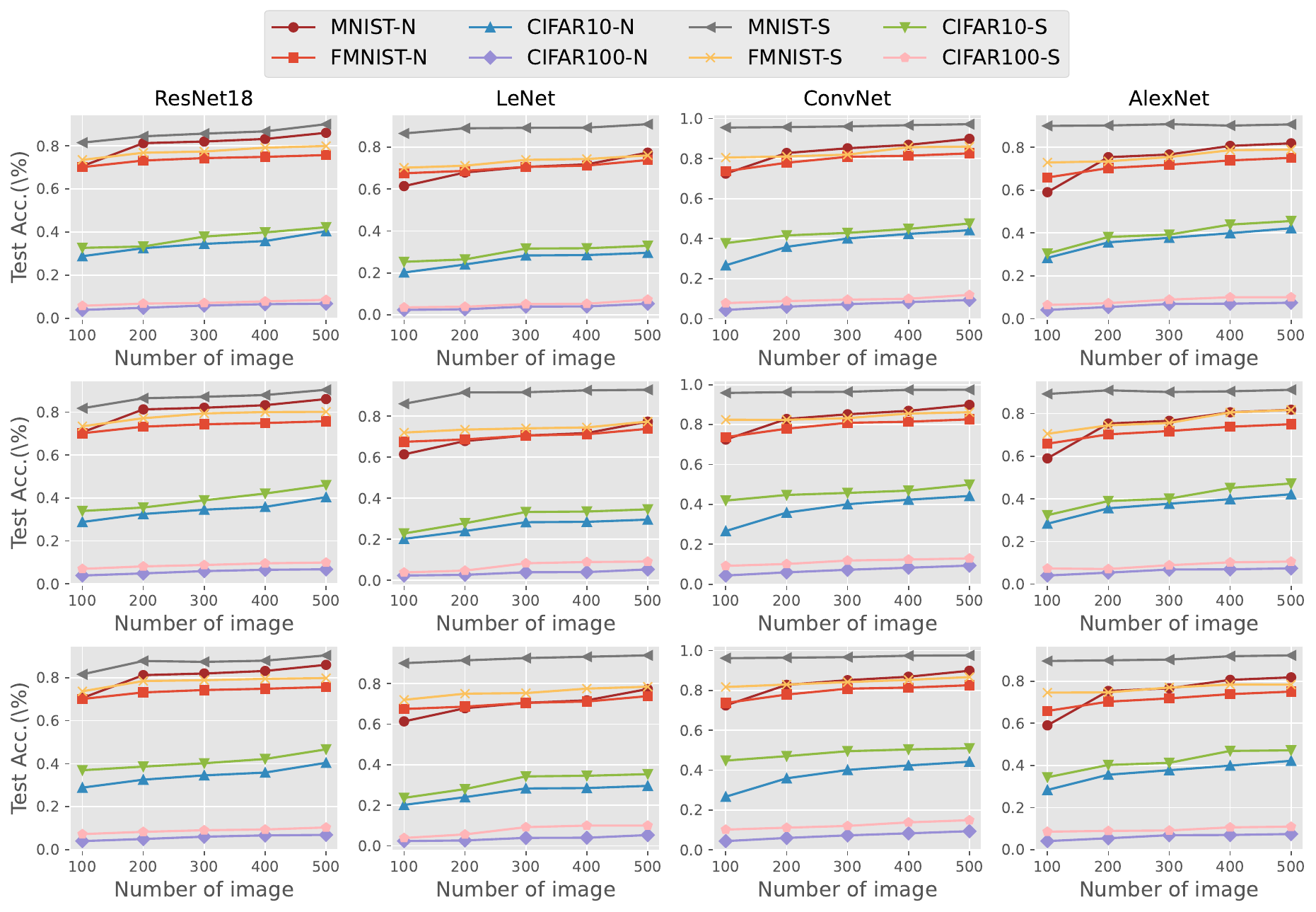}
	\caption{In scenario 2, the information varies between informative data (S) and normal data (N) across different networks as follows: Row 1 shows 1\%, 1\%, 2\%, and 2\% Dst. Kwl. for MNIST, FMNIST, CIFAR10, and CIFAR100, respectively; Row 2 shows 5\% Dst. Kwl.; Row 3 shows 10\% Dst. Kwl. We train both data types, S and N, on separate networks and then compare their accuracies to assess their relative information.}
	\label{scenario2info}
\end{figure}

\begin{figure}[tb]
	\centering
	\includegraphics[width=1\columnwidth]{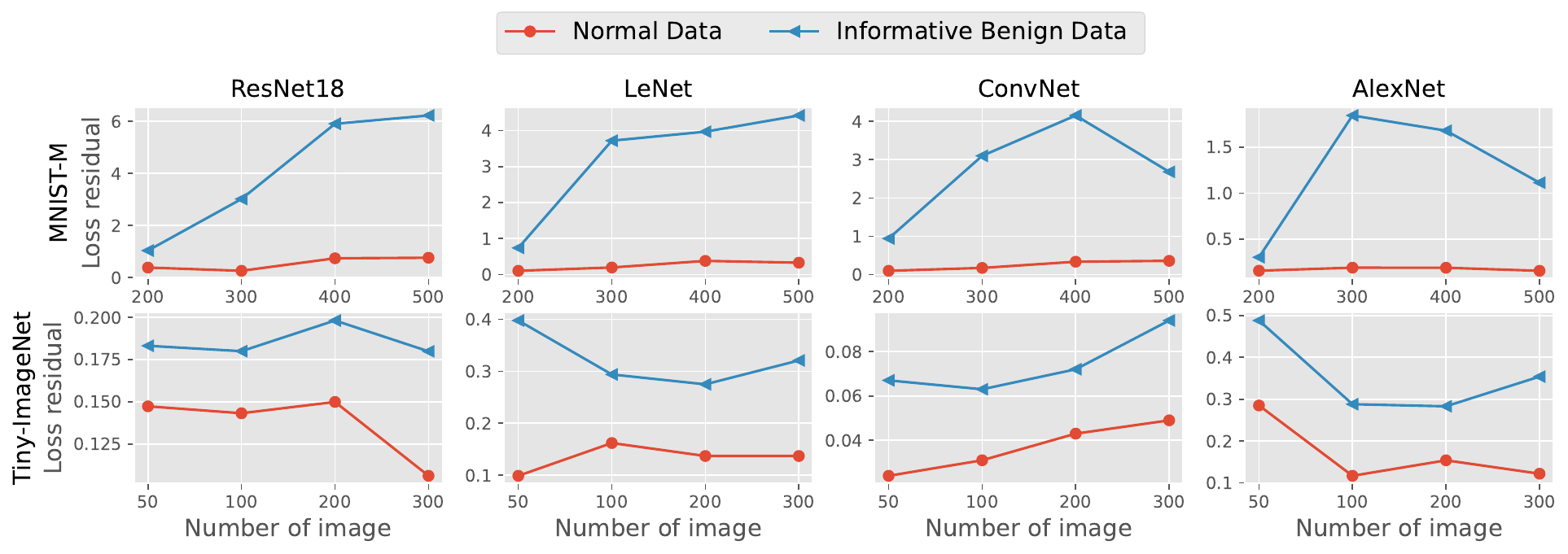}
	\caption{Comparison of mean loss discrepancy between informative and normal data on untrained and trained Networks in scenario 3.}
	\label{scenario3info}
\end{figure}

\textit{General Analysis: Information of Informative and Normal Data.}
We hypothesize that unlearning informative data (for all the evaluated unlearning methods) causes more degradation in model accuracy because it usually contains more information than normal data. 
To verify this hypothesis, we trained two separate networks on informative and normal data, respectively. We then compared their accuracy to assess the information of informative and normal data. The results for Scenario 1 are displayed in Figure \ref{scenario1info}, and the results for Scenario 2 results are illustrated in Figure \ref{scenario2info}. Our results across ResNet18, LeNet, ConvNet, and AlexNet demonstrated that the model trained on informative data consistently learn more information and thus achieve higher accuracy than on normal data in both scenarios. 
For Scenario 3, since the informative data comes from another domain, which is not the same as the domain of the normal data, we could not directly use the above method to verify the hypothesis. 
Instead, we trained multiple models on the datasets with or without the informative and normal data and computed the loss on these data. We compare the loss differences between including and excluding the informative or normal data in Figure \ref{scenario3info}, and we observed that the loss differences caused by including and excluding the informative data was significantly higher than the loss differences caused by including and excluding the normal data, indicating that the informative data contains more information from its domain. 

\textit{Additional Analysis for First-Order Method.} For the First-Order method \cite{r12}, which uses Formula \ref{eqcite12} for optimizing the \(\delta_x\), we further analyze the budget for unlearning Informative and Normal data. We chose to unlearn features and observe the differences in the required \(\delta_x\). To achieve flawless unlearning of informative and normal data, we must solve the required \(\delta_x\) using the specified Formula \ref{eqcite12}. Thus, to determine \(\delta_x\) and compare the budgets for unlearning normal and informative data, we exclude both types of data from the dataset. Subsequently, we retrain a model that has not been exposed to these datasets, denoted as \(\psi_{\theta_{us}}\).



We use \(\psi_{\theta_{us}}\) to compute \(loss_{nor}^{us}\) and \(loss_{inf}^{us}\) for normal and informative data, respectively. These losses reflect the performance of a model that has never been exposed to either type of data. We then train another model, \(\psi_{\theta_{se}}\), which has been exposed to both normal and informative data. Using this model, we add \( \delta_{nor} \) to normal data and \( \delta_{inf} \) to informative data. Through \( \psi_{\theta_{se}} \), we can obtain \( loss_{nor}^{se} \) and \( loss_{inf}^{se} \), which respectively represent the losses output by models with perturbed normal and informative data inputs.

First, we optimize \( \delta_{nor} \), the perturbation added to normal data, aiming to minimize the difference between \( loss_{nor}^{se} \) and \( loss_{nor}^{us} \). Similarly, we optimize \( \delta_{inf} \), the perturbation added to informative data, to minimize the difference between \( loss_{inf}^{se} \) and \( loss_{inf}^{us} \). Finally, we separately calculate the budgets required for the optimized \( \delta_{nor} \) and \( \delta_{inf} \).

We conducted experiments in three scenarios. Figure \ref{diffnoise} displays the optimal \(\delta_{inf}\) for informative data and \(\delta_{nor}\) for normal data. Specifically, the first and second rows of Figure \ref{diffnoise} show the optimal \(\delta_{inf}\) for scenarios 1 and 2, respectively. The third row presents the optimal \(\delta_{nor}\) across both scenarios. Additionally, the fourth row illustrates the optimal \(\delta_{inf}\) for informative data in scenario 3 using the MNIST dataset, while the fifth row depicts the optimal \(\delta_{nor}\) for normal data in the MNIST-M dataset.

\begin{figure}[tb]
	\centering
	\includegraphics[width=1\columnwidth]{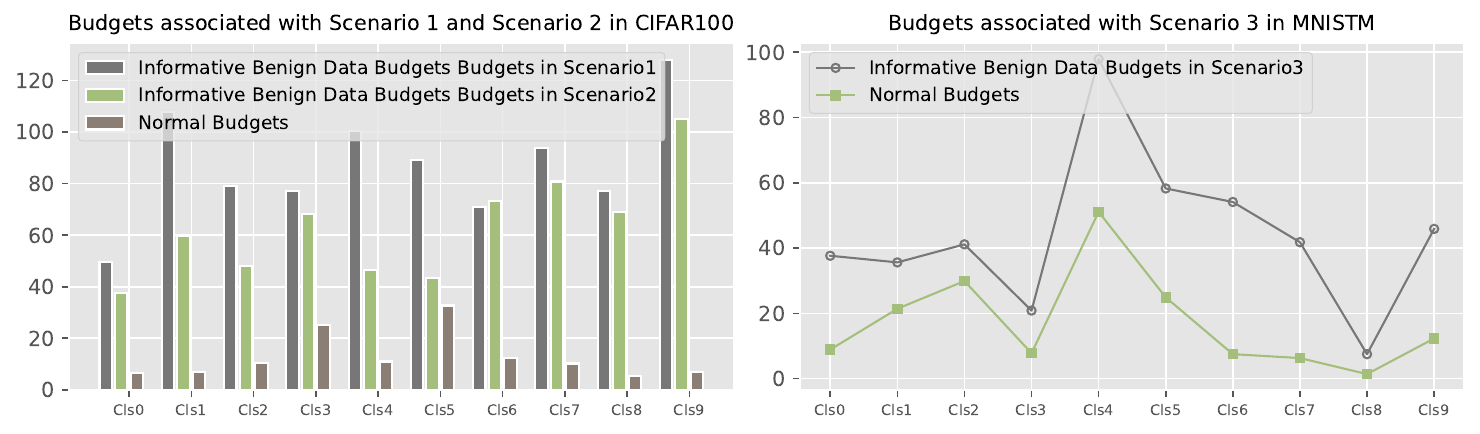}
	\caption{Comparison of the budget $\delta _{nor}$ required for unlearning normal data with the budget $\delta _{inf}$ needed for unlearning informative benign data.}
	\label{diffnoisebudget}
\end{figure}

The results demonstrate that unlearning informative data requires a larger budget, \(\delta_x\), due to the extensive range of image features it encompasses. In contrast, normal data necessitates a smaller \(\delta_x\), allowing the model to unlearn such images with a minimal budget. Figure \ref{diffnoisebudget} illustrates the budget disparities: the budget required for unlearning informative data is considerably higher than that for normal data, highlighting the challenges associated with completely unlearning informative data.

\textit{Additional Analysis for Amnesiac Unlearning.} For Amnesiac Unlearning, in the process of unlearning both normal and informative data, we meticulously analyzed their respective impacts on the model parameters. Amnesiac Unlearning provides a detailed account of the parameter updates influenced by both data types. Each type of data was processed in separate batches with no overlap, allowing us to isolate and record the updates induced solely by informative or normal data. Over 50 training epochs, we diligently documented these parameter updates for each batch.

During unlearning, parameters influenced by normal and informative data were selectively removed from the network. We employed Grad-CAM \cite{r34} to visualize the model's focus areas, with findings presented in Figure \ref{diffhot}. We tracked how the model's image recognition capabilities evolved during unlearning at epochs 0, 10, 20, 30, 40, and 50. Notably, as unlearning progressed—especially with informative data—the network's focus shifted from crucial recognition points to more dispersed areas. This pattern supports the observation that unlearning informative data removes a significant volume of the valuable contributions made by these parameters to the network, underscoring the positive impact of informative data on parameter updates compared to normal data.

\textit{In Appendix \ref{app:ablationoneimg}, we further utilize an extreme case study to unfold the impact of informative data on machine unlearning.}

\section{Resistance Against Poisoning Defenses}

In this section, we examine the effectiveness of poisoned data detection and defense mechanisms against informative benign data, as the success of such attacks hinges on the data's ability to bypass poisoned data detection tools in an automatic machine unlearning pipeline. Previous research has focused on detecting poisoned samples within datasets to protect models. If the prevailing automatic poisoning defensive mechanisms can detect and remove our informative samples, our attack strategy may fail. We evaluated the resilience of our informative data against both passive and active defenses on the CIFAR-10 dataset.

\textit{Resistance to Passive Defense.}  Tran et al. \cite{r39} observed that, while poisoned samples crafted by attacks like BadNets and Blend do not differ significantly from normal samples in the data space, their latent representations are mostly separable from the representations of normal samples. Based on this observation, Tran et al. \cite{r39} further proposed to use the top right singular value of the representation as a measure to separate the representations, which yields better results than using $\ell_2$ distance as a measure.

\begin{figure}[tb]
	\centering
	\includegraphics[width=1\columnwidth]{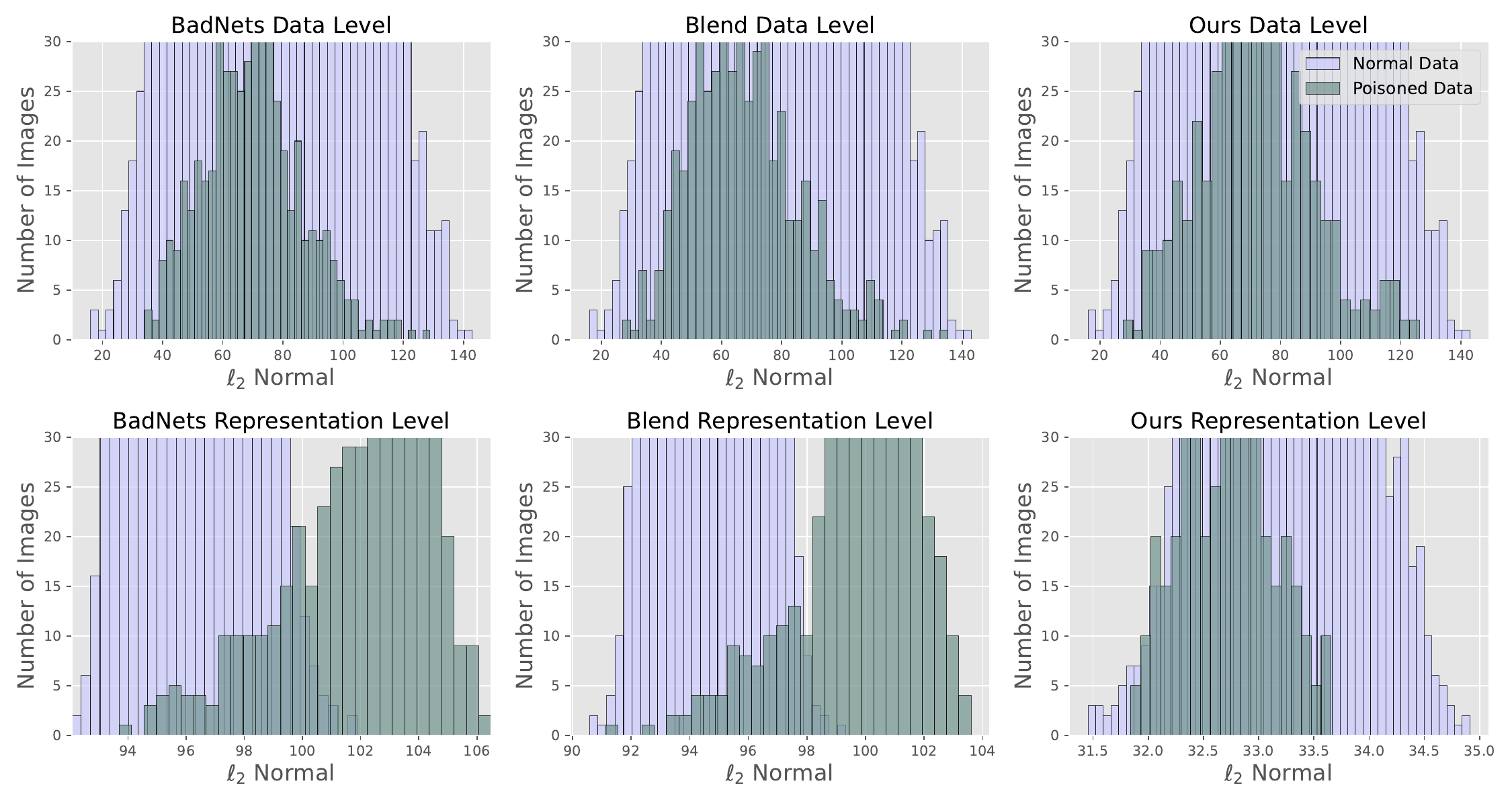}
	\caption{Comparative analysis of BadNets, Blend, and Informative Benign Data: separability in $\ell_{2}$ norm at data and representation levels.}
	\label{l2normalcompare}
\end{figure}

\begin{figure}[tb]
	\centering
	\includegraphics[width=1\columnwidth]{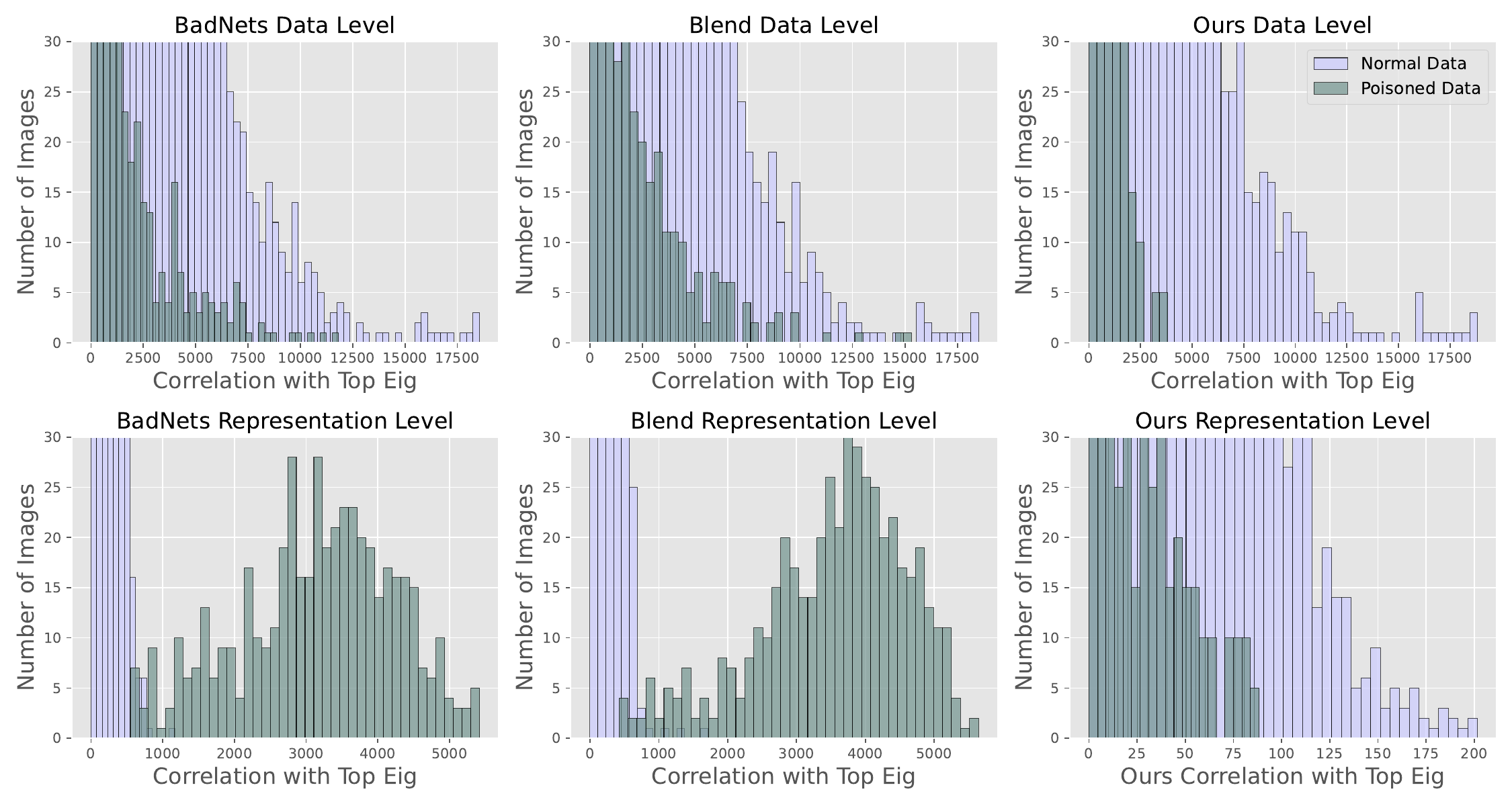}
	\caption{Comparative analysis of BadNets, Blend, and Informative Benign Data: separability in top eigenvalues at data and representation levels.}
	\label{topeigcompare}
\end{figure}

Here we assess if the informative benign data is separable from normal data in both data and latent spaces.
We show the results of using the $\ell_2$ distance as the measure in Figure \ref{l2normalcompare} and the results of using the top right singular value as a measure in Figure \ref{topeigcompare}.
The results showed that, our informative data does not exhibit separation from normal samples in both data and latent spaces under both of the measures. Therefore, Tran et al.'s method \cite{r39} can not detect our informative data, while it is very effective to detect poisoned data.

Hayase et al. \cite{r37} proposed a defense called SPECTRE that can effectively identify poisoned samples, even when the poisoning rate is low. SPECTRE estimates the mean and covariance of clean data and then whitens the data to align the initial PCA directions with the subspace differentiating poisoned from clean samples. Hayase et al. \cite{r37} further proposed the Quantum Entropy (QUE) outlier scoring method to improve separability between poisoned and normal samples.

\begin{figure}[tb]
	\centering
	\includegraphics[width=1\columnwidth]{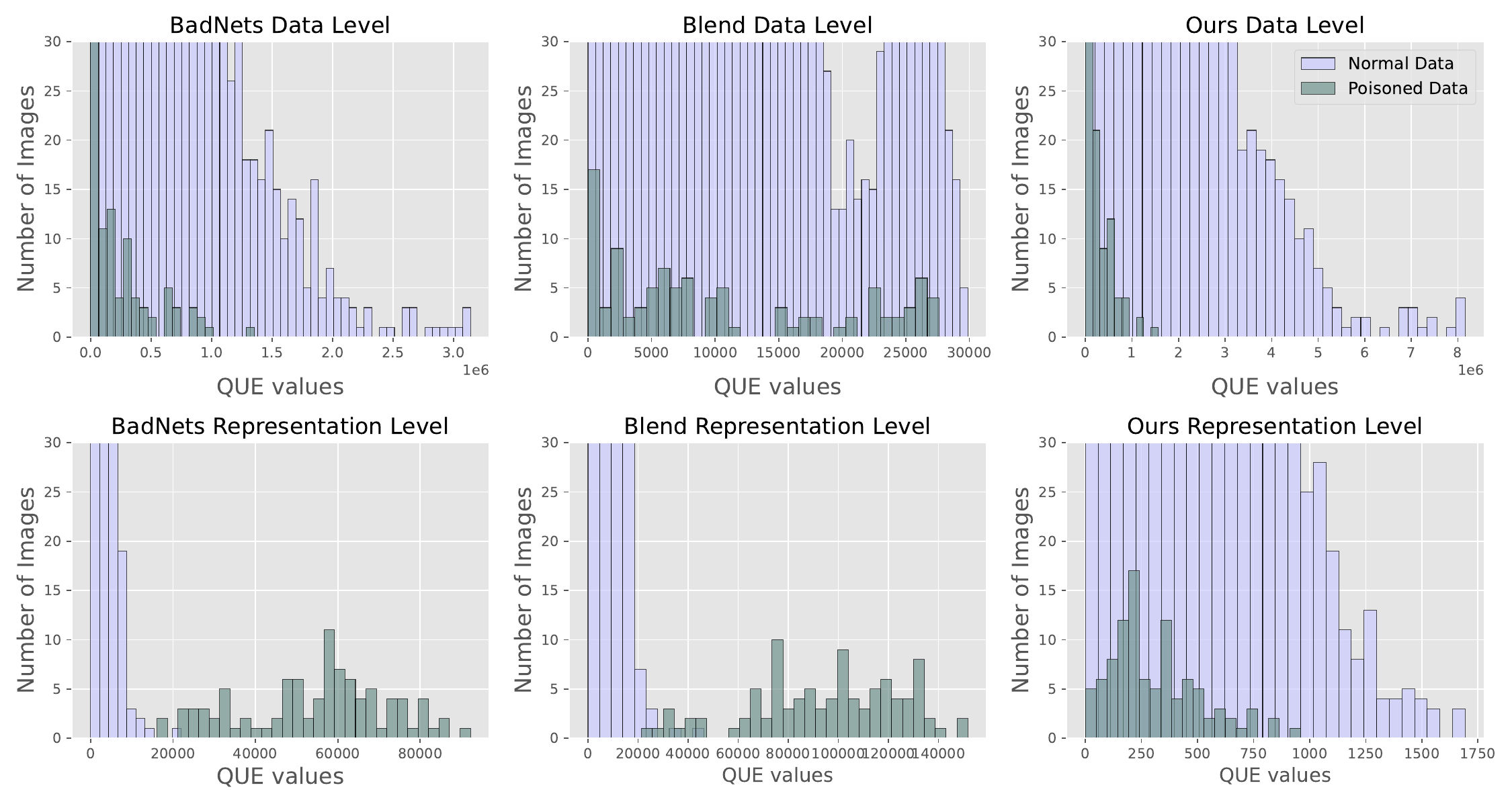}
	\caption{Comparative analysis of BadNets, Blend, and Informative Benign Data: separability in QUE values at data and representation levels.}
	\label{taucompare}
\end{figure}

We first calculate the QUE values for the poisoned samples crafted by BadNets and Blend and our informative samples in both the data and latent spaces to observe any separation phenomena. As shown in Figure \ref{taucompare}, even at low poisoning rates, the poisoned samples crafted by BadNets and Blend exhibit very high QUE values. Thus, poisoned data can be mostly removed by setting an appropriate threshold and removing all the data with high QUE values. 
In contrast, our informative samples exhibit low QUE values, which are similar to some clean samples. 


\begin{figure}[tb]
	\centering
	\includegraphics[width=1\columnwidth]{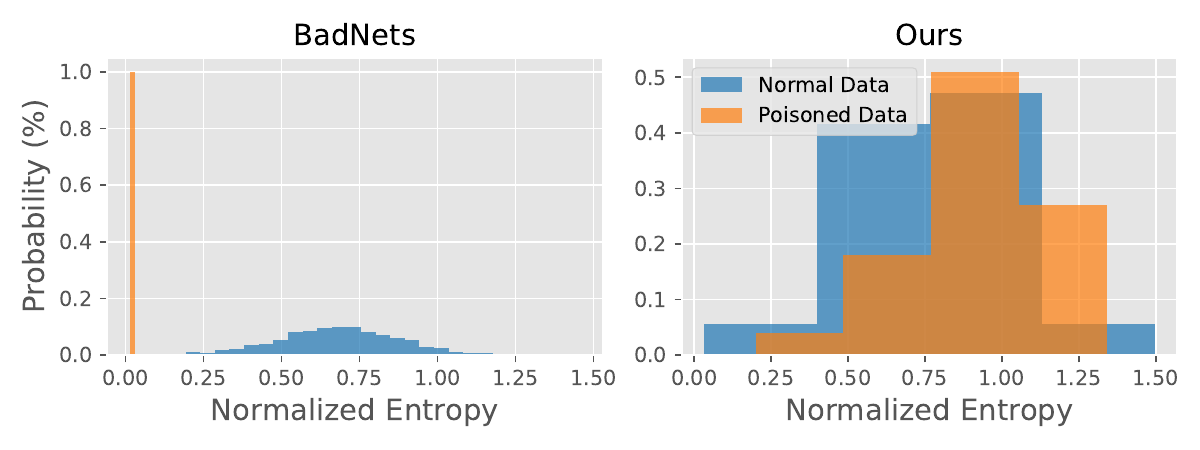}
	\caption{Comparing the resistance of informative benign data and BadNets against the Strip method on the CIFAR-10 dataset.}
	\label{stripcompare}
\end{figure}

Strip \cite{r41} creates perturbed images by overlaying training set images with clean ones and then uses a model to measure their entropy to detect poisoned data, as poisoned samples usually have lower entropy than normal ones. To evaluate the resistance of our informative data, we merged them with clean data and analyzed their entropy using the Strip method. As shown in Figure \ref{stripcompare}, unlike BadNets-generated samples, our informative samples maintain normal levels of entropy, indicating robustness against this detection method.

We show the detection results of four passive methods in Table \ref{passivedefense}, using TPR as an indicator to determine if our informative samples would be detected by these methods.
Table \ref{passivedefense} shows that all the evaluated passive defense methods are unable to detect our informative samples, which indicates that these samples are very likely to bypass a passive poisoned data detector in an automatic machine learning and unlearning pipeline.

\begin{table}[tb]
	\centering
	\caption{Comparative analysis of resistance to passive defenses: TPR evaluation on CIFAR-10.}
	\resizebox{0.48\textwidth}{!}{
		\begin{tabular}{cccccc}
			\hline
			
		   \textbf{Dataset} &\textbf{Method} & \textbf{BadNets} \cite{rbad}&\textbf{Blend} \cite{rblend}& \textbf{LC} \cite{rlc}& \textbf{Ours}   \\ \\[-2ex]\hline \\[-2ex]

            \multirow{5}{*}{Cifar10}& SS\cite{r39}& 100& 96.5& 57.5 &  \cellcolor{gray!30} 0      \\

            & SPECTRE\cite{r37}& 100 & 100& 99.6&    \cellcolor{gray!30} 0  \\
            & Strip\cite{r41}& 100& 89.2& 99&   \cellcolor{gray!30} 0   \\

            & Scan\cite{r38}& 100& 96.8& 97.8 &  \cellcolor{gray!30} 0  \\  \\[-2ex]\hline \\[-2ex]

		\end{tabular}
	}
	\label{passivedefense}
	
\end{table}

To further demonstrate how our informative benign data effectively bypasses passive defense mechanisms and to ensure the reliability of our findings, we conducted comprehensive experiments using both PCA and T-SNE dimensionality reduction techniques, known for their exceptional feature separation capabilities. 

\begin{figure}[tb]
	\centering
	\includegraphics[width=1\columnwidth]{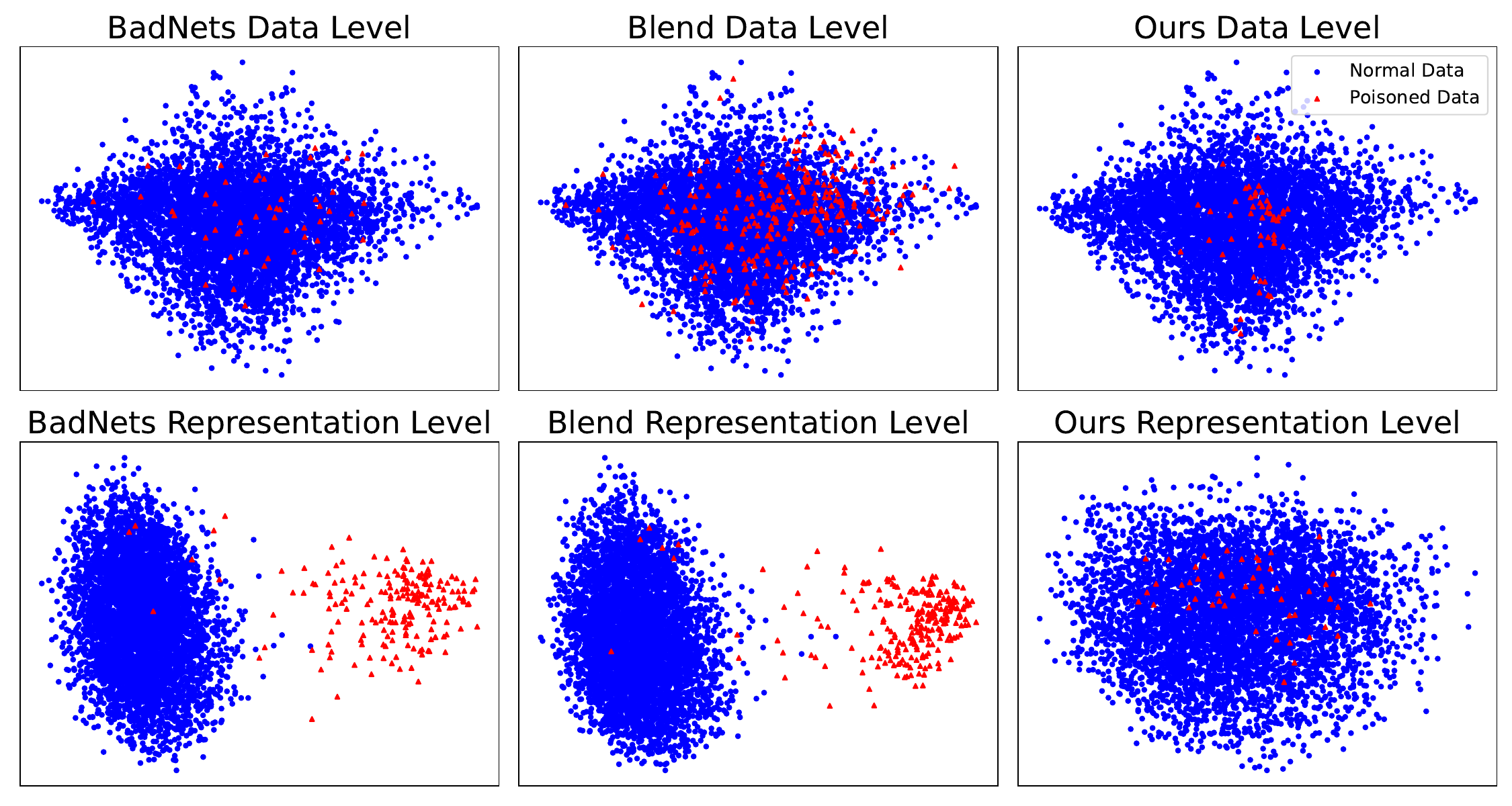}
	\caption{Comparative analysis of data distributions on CIFAR-10.}
	\label{pcacompare}
\end{figure}

We applied PCA dimensionality reduction \cite{rpca} to the CIFAR10 dataset, both at the data level and after feature extraction. Our results are illustrated in Figure \ref{pcacompare}. At the data level, our informative samples, along with those from BadNets \cite{rbad} and Blend \cite{rblend}, fall within the normal range of data distributions. PCA analysis helps visualize the distribution of data points in a lower-dimensional space, aiding in identifying patterns or anomalies.

However, after undergoing deep neural feature extraction, the distributions of the samples generated by BadNets and Blend significantly diverge from the clean data distribution. In contrast, our informative samples align closely with the distribution of clean data. 

\begin{figure}[tb]
	\centering
	\includegraphics[width=1\columnwidth]{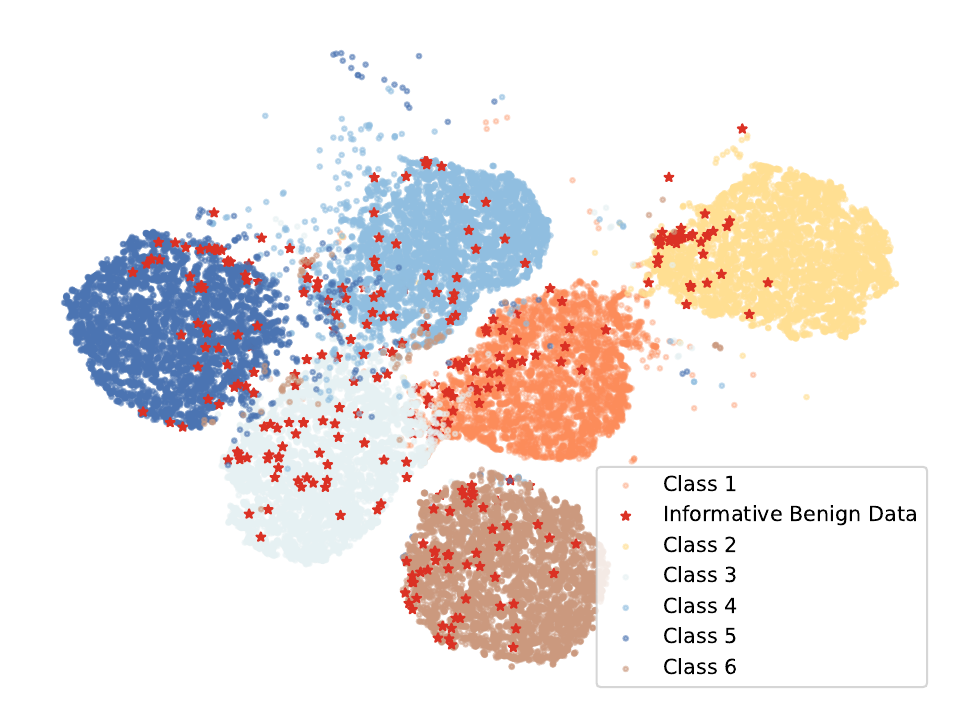}
	\caption{Analyzing feature distribution with T-SNE dimensionality reduction on CIFAR-10.}
	\label{tsnefeature}
\end{figure}

To further demonstrate the inseparability between our informative data and normal data, we employed the t-SNE \cite{rtsne} algorithm. We generated 50 informative samples for each class and injected them into the training dataset for network training. Subsequently, the network was utilized to extract features, enabling us to observe the feature distribution of our informative samples across six distinct categories, as illustrated in Figure \ref{tsnefeature}. Notably, unlike common poisoning samples, our samples' distribution does not exhibit any separation from the clean samples. This observation validates that the distribution of our generated informative benign data aligns with the norma data distribution, making it challenging for passive poisoned defense methods to detect.

\textit{Resistance to Active Defense.} Passive defense mechanisms utilize the inherent separability of the feature space to detect poisoned samples. However, Qi et al. \cite{qi2022revisiting} found that reducing the separability of attack samples in the feature space can improve their resistance to passive defenses. To overcome the limitation of passive defenses, Qi et al. \cite{r42} proposed an active defense method, which leverages label randomization to train a poisoned model for identifying poisoned samples and then employs clustering techniques to further cluster the poisoned samples for minimizing the false positive rate (FPR).

We evaluate whether our informative samples can bypass active defense, with the main results shown in Table \ref{activedefense}. We observed that both clean-label poisoned data \cite{rlc} and dirty-label poisoned data \cite{rblend, rbad} can be detected by active defense mechanisms. In contrast, most of the informative benign data can not be detected by these active defenses. This is because the informative data we generated is benign for machine learning, decoupled neural networks find it difficult to detect our informative benign data. Previous PCA and T-SNE algorithms also demonstrated that the data distribution of our informative benign data is similar to normal data distribution. Therefore, our research results indicate that, even if the automatic pipeline applies active defense strategies, detecting the informative benign data remains highly challenging. Consequently, our research motivates further work to rethink the concept of poisoned data and develop a defensive mechanism, as introduced below.


\begin{table}[tb]
	\centering
	\caption{Comparative analysis of active defense resistance: TPR evaluation on different data.}
	\resizebox{0.44\textwidth}{!}{
		\begin{tabular}{ccccc}
			\hline
			\makecell{\textbf{Method$\rightarrow$} \\ \textbf{Dataset$\downarrow$}} &\textbf{BadNets} \cite{rbad}& \textbf{Blend} \cite{rblend} & \textbf{LC} \cite{rlc}& \textbf{Ours} \\ \\[-2ex]\hline \\[-2ex]
			\textbf{MNIST} &100 & 100  &\diagbox{}{} & \cellcolor{gray!30}0 \\ 
			\textbf{FMNIST} &100 & 100 &\diagbox{}{}& \cellcolor{gray!30}0\\ 
			\textbf{CIFAR10} & 100& 100  &100& \cellcolor{gray!30}0 \\ \\[-2ex]\hline \\[-2ex]
			
		\end{tabular}
	}
	\label{activedefense}
	
\end{table}

\textit{Potential Defense.} To defend against our attack in an automatic machine unlearning pipeline, it is necessary to reconsider the definition of ``poisoned data'' within the context of machine unlearning. Although informative benign data is very similar to normal data in both data and latent spaces, it is typically more difficult to unlearn than normal data. Based on this observation, we can define the ``poisoned data'' as the data that is normal but costly to unlearn for machine unlearning.
Given this definition, we can identify data samples that require more computational cost and resources to unlearn or induce more loss change on themselves by monitoring the unlearning process. 
When a data sample’s unlearning cost is significantly higher than that of other data, we can trigger an alert mechanism that can remind the administrator to further investigate the sample offline. Additionally, we could maintain a validation dataset and evaluate the model performance after unlearning. If a significant degradation in the model performance is observed, the automatic pipeline could raise an alarm.
We hope future work can continue to explore the risk of informative data and develop defenses to enhance the security and stability of an automatic machine unlearning system.


\section{Conclusion}
This paper introduces the concept of the "unlearning usability attack" as a novel approach to overcoming the limitations present in current unlearning attack strategies. Our method differs from conventional approaches by avoiding the use of perturbed samples in unlearning requests, thus enabling attacks on automated machine unlearning systems while circumventing hash-based checks. We present three plausible attack scenarios: User-Developer Collusion, User Collusion, and Independent Users. Through these scenarios, our study illustrates that even within black-box environments, we can induce significant unlearning in models by leveraging Informative Benign Data. Additionally, our generated informative benign data demonstrates robust resistance against both passive and active defense mechanisms, prompting a reassessment of the conventional understanding of "poisoned data" in machine unlearning.

Given the increasing importance of machine unlearning services amid escalating privacy concerns, our study highlights the inherent vulnerabilities of these techniques within MLaaS frameworks, which lays the groundwork for understanding and mitigating these risks. Future research may focus on developing more robust unlearning procedures and enhancing data detection capabilities to maintain a balance between data privacy, model functionality, and security in MLaaS environments.






%
\bibliographystyle{IEEEtran}
\bibliography{IEEEabrv,bare_conf_compsoc}

\appendices

\section{Ablation Studies}\label{app:abstud}

\subsection{Different Amounts of Informative Benign Data in Scenario 1.}\label{app:absce1} 
\begin{figure*}[tb]
	\centering
	\includegraphics[width=1.7\columnwidth]{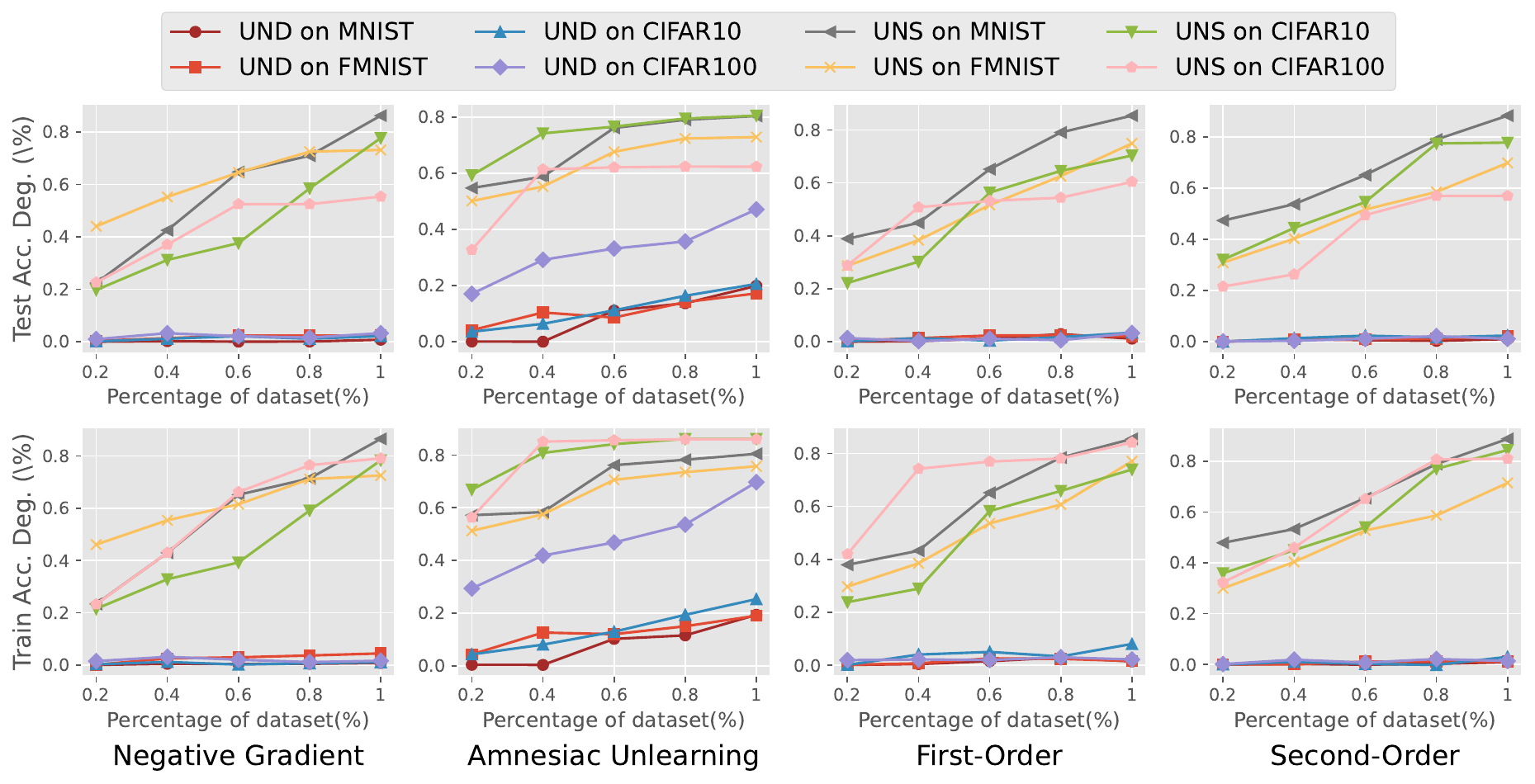}
	\caption{Comparison of accuracy change in the model after unlearning varying amounts of informative benign data and normal data, respectively.}
	\label{figdiffpoison}
\end{figure*}

By maintaining injection rates below 1\%, we observed that increasing these rates had minimal to negligible impact on the network's original accuracy; in some instances, accuracy even improved, which is less likely to cause concern in collaborative settings. Furthermore, as the injection rate increased, the decline in post-unlearning accuracy in the network became more pronounced compared to lower rates. This phenomenon is illustrated in Figure \ref{figdiffpoison}. This trend suggests that higher injection rates more effectively encapsulate users' data information within the informative benign samples, resulting in a greater impact on model performance after unlearning. 

\subsection{Different Network Architectures.}\label{app:abdiffnet} We use various network architectures to generate informative benign data as our informative dataset, aiming to explore whether these data are effective when attacking different network architectures. Through a series of experiments, attacks were executed on multiple network structures to assess the universality of our approach. We employed ConvNet \cite{r23}, AlexNet \cite{r35}, and ResNet18 \cite{r33} architectures to generate informative data, attacking ConvNet, AlexNet, and ResNet18, thereby examining the adaptability of our methodology. The outcomes resulting from attacks on diverse network architectures using informative benign data generated by distinct network architectures are presented in Table \ref{diffarch}. We observe that informative benign data generated with different network architectures is equally effective when targeting various network structures. However, informative data generated by ConvNet often exhibits better attack performance compared to those generated by other network architectures. This is attributed to the higher information content within informative data produced by ConvNet. In comparison to normal data, these informative samples contain more information, thus confirming the efficacy of our approach.

\begin{table}[tb]
	\centering
	\caption{In 0.2\% injection rate, the effectiveness of attacking different network architectures (measured by accuracy degradation (\%)) using informative benign data generated from various architectures is observed in the CIFAR-10 dataset.}
	\resizebox{0.48\textwidth}{!}{
		\begin{tabular}{ccccc}
			\hline
			\textbf{Method} & \diagbox{\textbf{Tar}}{\textbf{Sur}} &\textbf{ConvNet}&\textbf{AlexNet}&\textbf{ResNet18}\\ \hline
			
			\multirow{3}{*}{ \makecell{\textbf{Amnesiac}\\\textbf{Unlearning} \cite{r10}}} & \textbf{ConvNet} & \cellcolor{gray!30} \textbf{72.2}& \cellcolor{gray!10} 69.3 &\cellcolor{gray!20} 69.8 \\
			&\textbf{ AlexNet} &\cellcolor{gray!30} \textbf{68.2} & \cellcolor{gray!20} 67.8 & \cellcolor{gray!10} 67.5\\
			& \textbf{ResNet18} & \cellcolor{gray!20} 68.6 & \cellcolor{gray!10} 65.4 &\cellcolor{gray!30} \textbf{69.3 }\\ \hline
			
			\multirow{3}{*}{ \makecell{\textbf{Neg.}\\\textbf{Grad}}} & \textbf{ConvNet} &\cellcolor{gray!30}  \textbf{40.8} & \cellcolor{gray!10} 38.3 & \cellcolor{gray!20} 38.8 \\
			& \textbf{AlexNet} &\cellcolor{gray!30} \textbf{35.9} & \cellcolor{gray!20} 31.8 & \cellcolor{gray!10}31.4\\
			& \textbf{ResNet18} & \cellcolor{gray!30} \textbf{36.2} & \cellcolor{gray!20} 34.3 & \cellcolor{gray!10} 34.1 \\ \hline
			
			\multirow{3}{*}{ \makecell{\textbf{First-Order} \cite{r12} }} & \textbf{ConvNet} & \cellcolor{gray!20} 26.9& \cellcolor{gray!10} 25.6 & \cellcolor{gray!30} \textbf{28.1} \\
			&\textbf{ AlexNet} &\cellcolor{gray!20} 24.1 & \cellcolor{gray!30}  \textbf{ 24.4} &\cellcolor{gray!10} 22.0\\
			& \textbf{ResNet18} & \cellcolor{gray!10} 23.6 &\cellcolor{gray!20} 24.2 & \cellcolor{gray!30} \textbf{27.1} \\ \hline
			
			\multirow{3}{*}{ \makecell{\textbf{Second-Order} \cite{r12} }} & \textbf{ConvNet} & \cellcolor{gray!30}  \textbf{35.9}&\cellcolor{gray!10} 30.9 &\cellcolor{gray!20} 32.9 \\
			& \textbf{AlexNet} &\cellcolor{gray!20} 30.5 & \cellcolor{gray!30} \textbf{31.2} &\cellcolor{gray!10} 29.7\\
			& \textbf{ResNet18} & \cellcolor{gray!10} 33.5 &\cellcolor{gray!20}  34.1 & \cellcolor{gray!30} \textbf{35.2} \\ \hline
		\end{tabular}
	}
	
	\label{diffarch}
\end{table}

\subsection{Different Amounts of Informative Benign Data in Scenario 2.}\label{app:absce2} 
\begin{figure*}[tb]
	\centering
	\includegraphics[width=1.7\columnwidth]{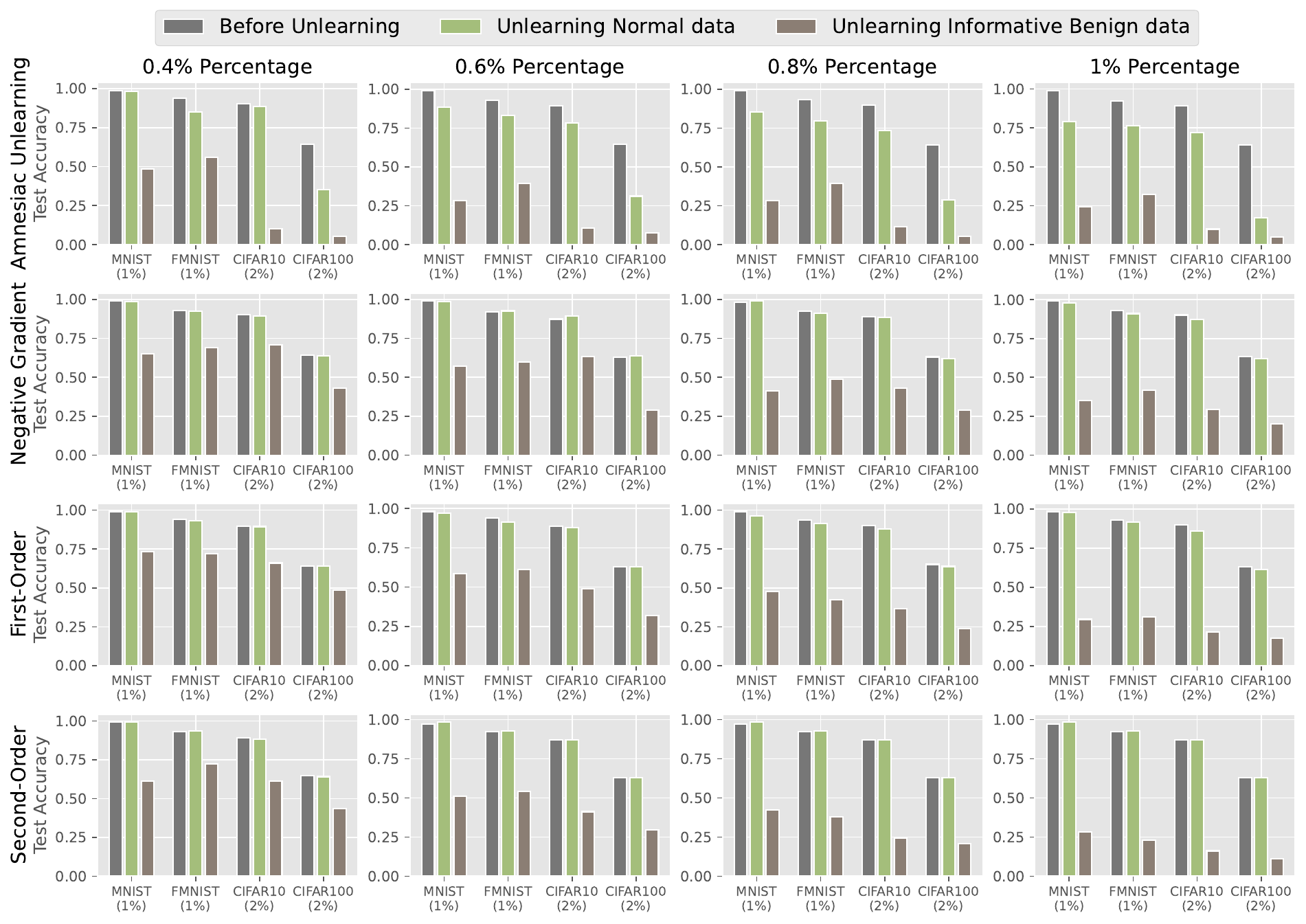}
	\caption{The impact of model accuracy after unlearning informative data with different injection rates in Scenario 2. Where "Before Unlearning" highlights the impact of informative data on initial model accuracy.}
	\label{diff_sce_poi_ra}
\end{figure*}

\begin{figure*}[tb]
	\centering
	\includegraphics[width=1.7\columnwidth]{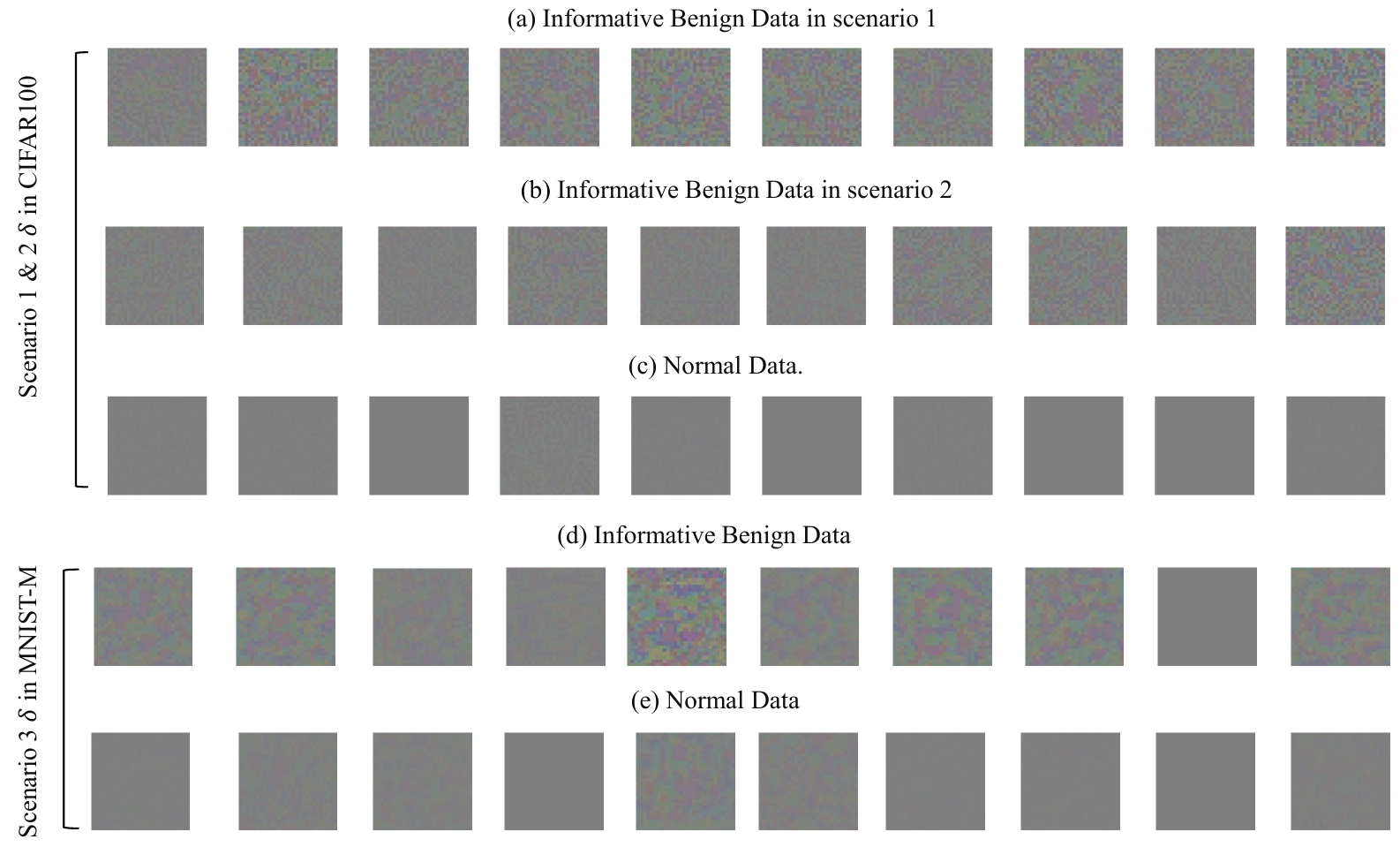}
	\caption{Comparison of the $\delta _{x}$ required for unlearning normal data and informative data. The first and second rows depict the $\delta _{inf}$ needed for unlearning informative benign data, and the third row illustrates the $\delta _{nor}$ needed for unlearning normal data in scenarios 1 and 2. The fourth and fifth rows illustrate the $\delta _{inf}$ and $\delta _{nor}$ needed for unlearning informative data and normal data in scenario 3, respectively.}
	\label{diffnoise}
\end{figure*}

\begin{figure*}[tb]
	\centering
	\includegraphics[width=1.7\columnwidth]{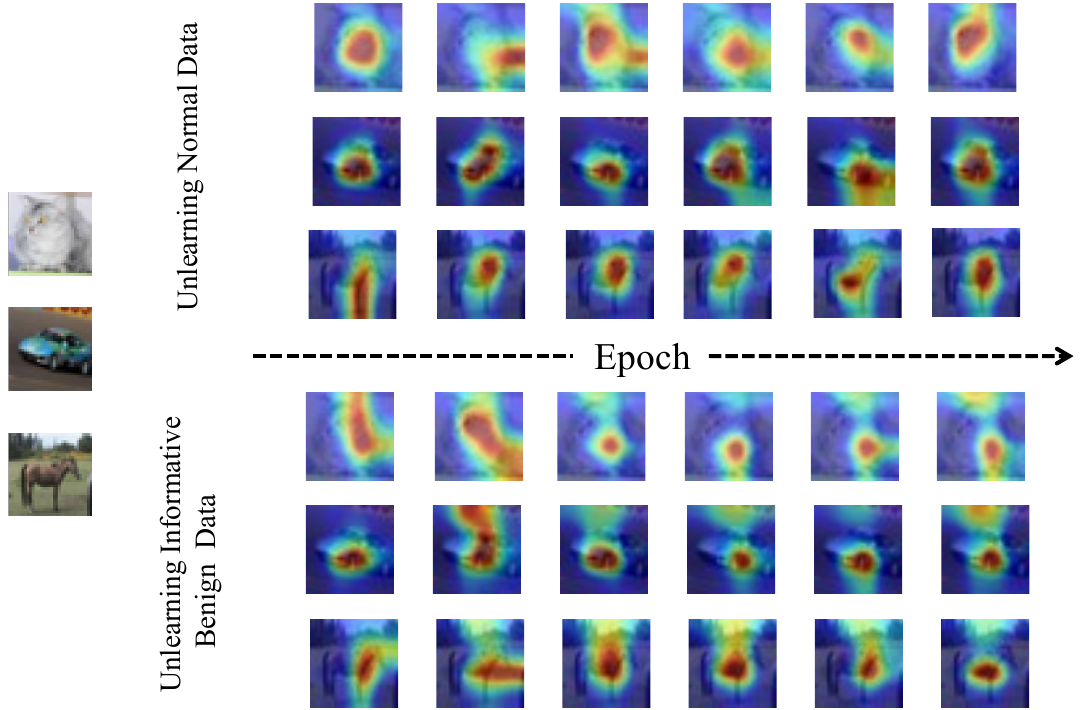}
	\caption{Comparison of how the attention of a model evolves during unlearning epochs for both normal data and informative data.}
	\label{diffhot}
\end{figure*}

Consistent with the findings in Scenario 1, as shown in Figure \ref{diff_sce_poi_ra}, we synthesized data using 1, 1, 2, 2 Dst. Kwl. for MNIST, FMNIST, CIFAR10, and CIFAR100 datasets and gradually increased the injection rates. We observed that when the injection rate is below 1\%, the impact on the network's initial performance can be negligible. Moreover, unlearning more informative data leads to a more significant decline in network accuracy compared to unlearning fewer informative data. This further emphasizes that removing informative benign data reduces accuracy, even if these samples do not contain the full knowledge of the unused dataset.

\begin{table}[tb]
	\centering
	\caption{Train models using both normal and informative benign data separately, and then compare the accuracy degradation during the unlearning process. The deeper the \colorbox{gray!30}{gray color}, the greater the accuracy decline.}
	\resizebox{0.48\textwidth}{!}{
		\begin{tabular}{ccccccc}
			\hline
			
			\multirow{3}{*}{\textbf{Dataset}} &\multirow{3}{*}{\textbf{Method}} & \multicolumn{2}{c}{Normal Data}& & \multicolumn{2}{c}{Informative Benign Data}   \\ \cline{3-4} \cline{6-7}
			
			& & \makecell{Before\\Unlearning}& \makecell{Acc.\\Deg.(\%)} & & \makecell{Before\\Unlearning}&  \makecell{Acc.\\Deg.(\%)} \\ \\[-2ex]\hline \\[-2ex]
			
			\multirow{4}{*}{ \makecell{MNIST\\(1Img/Cls)}} &\makecell{\textbf{Amn. Unl.}} & 48.9&  \cellcolor{gray!10} 48.7 & & 85.7& \cellcolor{gray!30} 84.9\\
			&\makecell{\textbf{Neg.}\textbf{Grad.}} &48.9 & \cellcolor{gray!10} 45.3 & &85.5 & \cellcolor{gray!30} 81.4 \\
			& \textbf{First-Order} & 49.0 & \cellcolor{gray!10} 48.1 & &85.5  & \cellcolor{gray!30} 83.8\\ 
			& \textbf{Second-Order} & 48.7 & \cellcolor{gray!10} 47.5 & & 85.9 & \cellcolor{gray!30} 84.2\\ 
			\\ [-2ex] \hline \\[-2ex]
			\multirow{4}{*}{ \makecell{FMNIST\\(1Img/Cls)}} &\makecell{\textbf{Amn. Unl.}} &52.5 &\cellcolor{gray!10}  51.1& & 68.9 & \cellcolor{gray!30}  67.9 \\
			&\makecell{\textbf{Neg.} \textbf{Grad.}} & 52.4& \cellcolor{gray!10} 46.2& & 68.9& \cellcolor{gray!30} 64.3\\
			& \textbf{First-Order} & 52.4& \cellcolor{gray!10} 51.0 & & 68.8 & \cellcolor{gray!30} 67.1\\ 
			& \textbf{Second-Order} & 52.2 & \cellcolor{gray!10} 49.4 &  & 69.0 &  \cellcolor{gray!30} 68.0\\ 
			\\ [-2ex] \hline \\[-2ex]
			\multirow{4}{*}{ \makecell{Cifar10\\(1Img/Cls)}} &\makecell{\textbf{Amn. Unl.}} & 13.0& \cellcolor{gray!10} 12.2&  & 28.2&  \cellcolor{gray!30} 27.4 \\
			&\makecell{\textbf{Neg.}\textbf{Grad.}} &13.1 & \cellcolor{gray!10} 10.8 & &28.1 &  \cellcolor{gray!30} 26.3\\
			& \textbf{First-Order} & 13.1 & \cellcolor{gray!10} 11.7 & & 28.2 &  \cellcolor{gray!30} 27.1 \\ 
			& \textbf{Second-Order} & 12.8 & \cellcolor{gray!10} 12.0& &  28.1& \cellcolor{gray!30} 26.8\\ 
			\\ [-2ex] \hline \\[-2ex]
			\multirow{4}{*}{ \makecell{Cifar100\\(1Img/Cls)}} &\makecell{\textbf{Amn. Unl.}} & 4.4 & \cellcolor{gray!10} 3.2& & 12 &  \cellcolor{gray!30} 11.6\\
			&\makecell{\textbf{Neg.}\textbf{Grad.}} & 4.4& \cellcolor{gray!10} 2.8& &  11.8&  \cellcolor{gray!30} 11.3 \\
			& \textbf{First-Order} & 4.4 & \cellcolor{gray!10} 3.8 & &  12 & \cellcolor{gray!30}  11.3\\ 
			& \textbf{Second-Order} & 4.4 & \cellcolor{gray!10} 4.1&  &  12 &  \cellcolor{gray!30} 11.8\\ \hline
		\end{tabular}
	}
	\label{onlypoi}
\end{table}

\section{The impact of two types of one image on unlearning}\label{app:ablationoneimg}

\noindent\textbf{The effect of separately training two networks with one normal and one informative benign data from each class on machine unlearning.} We trained the model separately using real data and informative data. Subsequently, we applied the unlearning method to individually unlearn real data and informative data, observing the impact of these two types of data on the model's accuracy. For our informative data, we generated informative benign data using the Convolutional (Conv) architecture. Additionally, we created one informative image (1Img/Cls) for each class as training data. Simultaneously, we randomly selected one image (1Img/Cls) from the dataset for each class as training data. Following this, we conducted separate training sessions for these two sets of training data using the Conv architecture. Finally, we examined the extent of accuracy degradation in the model after unlearning.

In this extreme scenario, we compared the model's accuracy decline, demonstrating the effectiveness of our attack. Unlearning our informative data resulted in a more substantial accuracy drop, as shown in Table \ref{onlypoi}. We observed that when training models separately with real data and informative data, the latter contained a greater amount of information. Consequently, during the training of models, informative data achieved a higher network accuracy compared to real data. Moreover, we found that unlearning a network trained solely on informative data resulted in a more significant accuracy drop compared to unlearning a network trained on real data. When unlearning real data, the network's accuracy decline was smaller than in the case of unlearning informative data. Therefore, this ablation experiment intuitively demonstrates that unlearning informative data significantly impacts the network's accuracy.

\end{document}